\documentclass[fleqn,usenatbib]{mnras}

\usepackage{newtxtext,newtxmath}
\usepackage{mathptmx}
\usepackage{txfonts}
\usepackage[table]{xcolor}
\newcommand{\source}{{MAXI~J1820+070}}
\newcommand{\sourcem}{{MAXI~J1803-298}}
\newcommand{\sourceg}{{GX~339-4}}
\newcommand{\nicer}{\textit{NICER}}

\usepackage[T1]{fontenc}

\DeclareRobustCommand{\VAN}[3]{#2}
\let\VANthebibliography\thebibliography
\def\thebibliography{\DeclareRobustCommand{\VAN}[3]{##3}\VANthebibliography}

\usepackage{graphicx}	% Including figure files
\usepackage{amsmath}	% Advanced maths commands
\usepackage{amssymb}	% Extra maths symbols

\title[QPO lags and coherence in MAXI J1820+070]{Variable QPO lags and reduced coherence between the disc and corona in MAXI J1820+070}

\author[Bollemeijer \& Uttley]{
Niek Bollemeijer,$^{1}$\thanks{E-mail: n.a.bollemeijer@uva.nl}
Phil Uttley$^{1}$
\\
$^{1}$Anton Pannekoek Institute for Astronomy, Amsterdam, Science Park 904, NL-1098 NH, The Netherlands
}

\date{Accepted XXX. Received YYY; in original form ZZZ}

\pubyear{2025}

\begin{document}
\label{firstpage}
\pagerange{\pageref{firstpage}--\pageref{lastpage}}
\maketitle

\begin{abstract}
Quasi-periodic oscillations (QPOs) are observed in the hard state of many black hole X-ray binaries. Although their origin is unknown, they are strongly associated with the corona, of which the geometry is also subject to discussion. We present a thorough spectral-timing analysis of QPOs and broadband noise in the high-inclination BHXRB MAXI J1820+070, using the rich NICER data set of the source in the bright hard state of its outburst in 2018. We find that there is a large QPO hard lag between soft energy bands with significant disc emission and harder coronal power-law bands, which is absent when measuring lags between energy bands dominated by the coronal emission. The QPO lags between a soft band (with significant disc emission) and harder coronal power-law bands vary significantly with power-law flux, on time-scales of (tens of) seconds or a few QPO cycles, especially at QPO frequencies $\lesssim0.3$ Hz. At the same time, the QPO is found to be related to a decreased coherence between energy bands with significant disc emission and harder bands both at and below the QPO frequency, suggesting the QPO mechanism filters out part of the variability. Similar patterns in the frequency-dependent lags and coherence are observed in the BHXRB MAXI~J1803-298, which is a (dipping) high-inclination source, but not in the low-inclination source GX~339-4. We suggest that these findings may be evidence of changes in the vertical extent of the corona on time-scales slightly longer than the QPO cycle.
\end{abstract}


%%%%%%%%%%%%%%%%%%%%%%%%%%%%%%%%%%%%%%%%%%%%%%%%%%

%%%%%%%%%%%%%%%%% BODY OF PAPER %%%%%%%%%%%%%%%%%%

\section{Introduction}

Quasi-periodic oscillations (QPOs) are variations in X-ray flux on well-defined time-scales that have been observed in many black hole X-ray binaries (BHXRBs). They are observed over a wide range of X-ray energies \citep{Ma_2021} and despite their ubiquity, there is no consensus on their origin \citep{Ingram_2019review}. QPOs are strongly associated with a spectral component that looks like a power-law with a thermal cut-off at high energies, arising from a region close to the black hole that is known as the corona. Like the origin of QPOs, the nature and geometry of the corona remains the subject of an ongoing debate in BHXRB research and scenarios range from a hot inner flow \citep{Ferreira_2006,Veledina_2013hotflow,Marcel_2021,Kawamura_2022} to the base of the radio-emitting jet \citep{Markoff_2005,Kylafis_2008,Reig_2021}. The relative contribution of the accretion disc, which can dominate at softer X-rays, and the corona to the energy spectrum changes in a systematic way during a BHXRB outburst, leading to the definition of different accretion states \citep{Remillard_2006,Kalemci_2022}. 

QPOs are often observed during the hard state and the hard-intermediate state (HIMS). Although QPOs are mainly associated with the emission from the corona \citep{Sobolewska_2006}, studying the properties of QPOs at lower X-ray energies, where disk thermal emission is significant, may shed new light on both the geometry of the corona and on the origin of QPOs.

Phenomenologically, QPOs are divided into three types: A, B and C \citep{Casella_2005,Motta_2016}. Type-C QPOs are by far the most common and we refer to them as QPOs from now on. Different models for QPOs exist, which can be roughly divided into two broad classes \citep{Ingram_2019review}. In the past decade, it has been observed that several QPO properties, such as the rms-variability \citep{Motta_2015}, sign of the QPO lags \citep{VandenEijnden_2017} and QPO waveform \citep{De_Ruiter_2019} depend on the inclination at which we view the binary system. Such a dependence is most naturally explained if the QPO is due to a \textit{geometric} change in the corona, for example due to (Lense-Thirring) precession \citep{Ingram_2009LT} or a precessing jet \citep{De_Ruiter_2019,Ma_2021}. However, other models try to explain QPOs through \textit{intrinsic} variations in flux due to instabilities such as the accretion-ejection instability \citep{Varniere_2002} or the jet instability \citep{Ferreira_2022}. The model \texttt{vKompth} \citep{Bellavita_2022} assumes a quasi-periodic change in coronal electron temperature due to heating variations and then enables fitting of the QPO lags by Compton delays in a number of sources, including \source{} \citep{MaRuican_2023}. 

Over a much broader range of frequencies, including the QPO-frequency, the X-ray light curves from BHXRBs in the hard state contain aperiodic variability. Such broadband noise variability is also associated with complex energy- and time-scale-dependent lags. Between a soft X-ray band with significant disc emission and harder bands, large hard lags of up to a second are measured at low frequencies, while at higher frequencies, the lags are often observed to be soft \citep{Wang_2022_revmachine}. Between higher energies, in power-law-dominated bands, the measured lags are hard.

Broadband noise is often modelled with accretion rate fluctuations propagating inward through the accretion disc \citep{Lyubarskii_1997,Churazov_2001}. In such a framework, hard emission from the corona follows soft emission from the disc (leading to hard lags), as is observed in many sources \citep{Arevalo_2006,Ingram_2011,Uttley_2011,Uttley_2014review}. 
The hard lags between harder X-ray bands have been modelled in several ways. For example, the hard lags have been attributed to Comptonization delays in a large corona \citep{Kazanas_1997,Reig_2003,Bellavita_2022}, while they can also be explained by a propagating fluctuations in a stratified corona with harder emission originating from closer to the black hole \citep{Kotov_2001,Mahmoud_2018,Kawamura_2022}. More recently, \citet{Uttley_2025} showed that variations in heating and seed photon flux to the corona (due to the propagating fluctuations) can cause the power-law spectrum to pivot, creating delays between power-law-dominated energy bands.

In many sources, there is clear evidence of reprocessing of coronal emission by the accretion disk, especially at the Fe K line around 6.4 keV and the Compton hump at higher energies, which are collectively known as the reflection spectrum (e.g. \citealt{Garcia_2014,Bambi_2021review}). In some models, reflected emission from the corona is used to explain the high-frequency soft lags between soft X-rays (with a significant contribution from the accretion disc) and harder bands \citep{Kara_2019}. In such models, lags arise due to reverberation of coronal emission on the disc. The reprocessed photons have a longer path length than the direct emission from the corona, leading to lags \citep{Uttley_2011,Ingram_2019reltrans,Mastroserio_2021}. 

In \citet{Bollemeijer_2024}, we found that the high-frequency lags that are associated with reverberation change on a broad range of time-scales. The lags were found to be strongly correlated with the instantaneous flux from the corona and the strongest relation between flux and lags was measured for variability at the QPO time-scale. Since reverberation lags are thought to depend on the coronal geometry \citep{Kara_2019,Uttley_2025}, we interpret those findings as evidence for a dynamic corona that varies in geometry, especially at the QPO time-scale. In the current work, we extend the analysis in \citet{Bollemeijer_2024} by measuring the lags at the QPO frequency itself, and we study how the QPO lags depend on the hard X-ray flux on longer time-scales. If the coronal geometry varies on both the QPO- and other time-scales, we expect that the QPO lags themselves will change as well. In several QPO models, the QPO lags depend on the geometry of the corona \citep{Stevens_2016,Bellavita_2022}, and in this work, we investigate whether the QPO lags depend on variations of source flux on time-scales of a few QPO cycles, i.e. seconds to minutes.

Time-scale-dependent variability can be well-studied in the Fourier domain. In this work, we focus on the cross-spectral properties of different energy bands, in particular on the coherence and phase lags \citep{Vaughan_1997coherence,Uttley_2014review}. The cross-spectrum of light curves from two different energy bands contains information on how variability from separate spectral components are related. The cross-spectral phase lags allow us to study time-scale-dependent delays between variability in two energy bands. The frequency-dependent coherence function measures to what extent variability in two light curves is linearly correlated and is defined to be between 0 and 1. Non-unity coherence implies that there are separate sources of variability in the light curves or that they are related in non-linear ways \citep{Vaughan_1997coherence,Nowak_1999}. 
\newline

\source{} (ASASSN-18ey, \citealt{Torres_2019}) is a low mass X-ray binary with a black hole mass of $5.73 < M_{BH} \left( M_{\odot} \right) < 8.34 $ and viewed at a high inclination of $>63^{\circ}$ \citep{Torres_2020}. During its outburst in 2018, it was observed extensively by several X-ray telescopes, which led to many spectroscopic and timing analyses (e.g. \citealt{Buisson_2019,Homan_2020,You_2021,Wang_2021_J1820,Kawamura_2022,Fan_2024}). Its low interstellar absorption ($\rm{N_H}=\sim1.5\times10^{21} \rm{cm}^{-2}$, \citealt{Uttley_2018}) enables study of soft X-rays down to 0.3 keV with \nicer. \source{} remained in the hard state for several months before transitioning to the soft state, providing an unprecedented view of the X-ray variability at low X-ray energies. We analyse selected \nicer{} observations, which are listed in Table \ref{tab:obse_overview}.

The paper is set up as follows. In section \ref{sec:obs}, we describe how we obtained, reprocessed and selected the \nicer{} observations used. In section \ref{sec:analysis_results}, we describe our analysis and results, which can be divided into three parts. First, we extend the analysis presented in \citet{Bollemeijer_2024} by investigating the flux-dependence of the QPO lags in \source{} on time-scales slightly longer than the QPO time-scale. In subsection \ref{subsec:engdep_lags_coh}, we study the energy-dependence of the lags and coherence, showing how these properties are related to the QPO. Finally, we compare our results for \source{} to data from BHXRBs \sourcem{} and \sourceg{} in subsection \ref{subsec:1803_lagcoh}. In the Discussion, we suggest a possible mechanism to explain both the variable QPO lag and the decreased disc--power-law coherence at and below the QPO time-scale, which may be related to variations in coronal geometry (section \ref{sec:discussion}). 

\begin{table*}
    \centering
    \begin{tabular}{cccccccc}
        ObsID & Date & MJD & $\nu_{\rm{QPO}}$ (Hz) & Hue (deg) & HR (4-10 / 2-4 keV) & 0.5-10 keV flux (cts/s)\\
        \hline\hline
        \textbf{\source} &&\\
        \hline
        
        1200120104& 15-03-2018 & 58192 & - & -12.1 $\pm$ 1.5 & 0.3488 $\pm$ 0.0004 & 3106 \\
        \rowcolor{gray!30}
        1200120108& 23-03-2018 & 58200 &0.034 & 1.8 $\pm$ 1.9 & 0.3145 $\pm$ 0.0003 & 19160 \\
        1200120112& 27-03-2018 & 58204  &0.044& 11.7 $\pm$ 1.5 & 0.312 $\pm$ 0.0002 & 18165\\
        \rowcolor{gray!30}
        1200120120& 04-04-2018 & 58212 & 0.066& 28.6 $\pm$ 2.9 & 0.3097 $\pm$ 0.0002 & 17307 \\
        1200120130& 16-04-2018 & 58224 &0.12 & 63.8 $\pm$ 2.9 & 0.3049 $\pm$ 0.0003 & 16822\\
        1200120131& 17-04-2018 & 58225 &0.12 & 64 $\pm$ 4 & 0.3044 $\pm$ 0.0003 & 17530\\
        \rowcolor{gray!30}
        1200120135& 23-04-2018 & 58231 &0.17 & 90.2 $\pm$ 2.8 & 0.3004 $\pm$ 0.0003 & 16981\\
        \rowcolor{gray!30}
        1200120136& 24-04-2018 & 58232 &0.18 & 69 $\pm$ 13 & 0.299 $\pm$ 0.0007 & 16769\\
        \rowcolor{gray!30}
        1200120137& 25-04-2018 & 58233 &0.18 & 86.2 $\pm$ 2.8 & 0.3002 $\pm$ 0.0003 & 16520\\
        1200120141& 01-05-2018 & 58239 & 0.28 & 120 $\pm$ 4 & 0.2941 $\pm$ 0.0005 & 16402\\
        1200120142& 02-05-2018 & 58240 & 0.28 & 117.3 $\pm$ 1.4 & 0.2927 $\pm$ 0.0002 & 16188\\
        1200120143& 03-05-2018 & 58241 & 0.30 & 121.5 $\pm$ 1.4 & 0.2919 $\pm$ 0.0003 & 16118\\
        \rowcolor{gray!30}
        1200120145& 05-05-2018 & 58243 & 0.34  & 129.1 $\pm$ 1.0 & 0.2906 $\pm$ 0.0002 & 16250\\
        1200120147& 07-05-2018 & 58245 & 0.38  & 132.0 $\pm$ 1.0 & 0.2886 $\pm$ 0.0002 & 16306\\
        1200120148& 08-05-2018 & 58246 & 0.40  & 135.7 $\pm$ 1.1 & 0.2873 $\pm$ 0.0002 & 16444\\
        \rowcolor{gray!30}
        1200120156& 22-05-2018 & 58260 & 0.50 & 138.2 $\pm$ 1.0 & 0.2831 $\pm$ 0.0002 & 14812\\
        1200120187& 24-06-2018 & 58293 & (0.28) & 110 $\pm$ 6 & 0.2981 $\pm$ 0.0007 & 6359 \\
        \hline
        \textbf{\sourceg} &&\\
        \hline
         3588011501 & 21-03-2021 & 59294 & 0.49 &  92.9 $\pm$ 2.9 & 0.3228 $\pm$ 0.0005 & 1259\\
        \hline
        \textbf{\sourcem} &&&&\\
        \hline
        \rowcolor{gray!30}
        4202130102 & 03-05-2021 & 59337 & 0.13 & 102 $\pm$ 15 & 0.3329 $\pm$ 0.0023 & 863 \\
        4202130104 & 05-05-2021 & 59339 & 0.36 & 110 $\pm$ 5 & 0.3340 $\pm$ 0.0008 & 1278 \\
        \hline
    \end{tabular}
    \caption{The full list of observations used for \source, \sourceg{} and \sourcem. For a more detailed account of power-spectral fits for \source, we refer to \citet{Stiele_2020}. The hue was calculated using the 4.8-9.6 keV energy band. The total flux is the count rate normalised to 52 functioning FPMs in \nicer. Consecutive lines with the same grey or white shading correspond to observations with very similar properties, which have been combined to increase the signal strength when flux binning. The brackets around the QPO frequency for observation 1200120187 refer to the fact that we do not observe a QPO in the power spectrum, but there are QPO-like features in the lags and coherence (see section \ref{sec:discussion}).}
    \label{tab:obse_overview}
\end{table*}

\begin{table}
\label{tab:band_names}
    \centering
    \begin{tabular}{c|c|c}
        \textbf{Energy range (keV)} & \textbf{Name} & \textbf{Spectral component} \\
        \hline\hline
        0.3-0.6 & very soft & disc \\
        0.6-1.3 & soft  & disc \\
        1.3-2   & medium-soft & power-law \\
        2-3     & medium-hard & power-law\\ 
        3-10    & hard  & power-law \\
    \end{tabular}
    \caption{The names used in this work for the different energy bands. The last column shows the spectral component associated with each energy band, which we use to refer to several energy bands collectively.}
\end{table}

\section{Observations}
\label{sec:obs}
We present a spectral-timing analysis of selected NICER observations of the 2018 outburst of BHXRB MAXI J1820+070 during the hard state. The observations of the hard state of \source{} were reprocessed using the \texttt{nicerl2} pipeline from HEASARC v6.33 \citep{FTOOLS_2014}, using default reprocessing settings. We selected 16 observations, which cover the full range of QPO frequencies in the hard state ($\sim$0.03-0.5 Hz). The high source brightness of \source{} led to telemetry issues in \nicer, which were solved by switching off a number of the 52 functioning Focal Plane Modules (FPMs). We made sure to only include FPMs that were switched on and passed the \texttt{nicerl2} screening criteria during the entire observation. When combining observations, we normalised the light curves by the number of FPMs used. In Table \ref{tab:obse_overview}, we show the combined observations by shading consecutive lines in the same colour, i.e. grey or white.

We also show spectral-timing results for \nicer{} observations of BHXRBs \sourcem{} and \sourceg{} in the hard state, both during their outbursts in 2021. The two observations for these sources were reprocessed in the same way as outlined before for \source. When comparing observations of different sources, we establish that they correspond to similar accretion states using two properties: the 4-10 / 2-4 keV hardness ratio (HR) and the power-spectral hue. The hue is based on power-colours (PC), which are ratios of the fractional rms$^2$ integrated over the following frequency ranges: PC1 is the variance in 0.25–2 Hz/0.0039–0.031 Hz, while PC2 is the variance in 0.031–0.25 Hz/2–16 Hz \citep{Heil_2015}. BHXRBs in outburst follow a wheel-shaped pattern when plotting PC1 versus PC2, which can be quantified using the hue. The hue is defined in degrees and ranges between 0 and 360$^{\circ}$, while the hard state is defined to lie between 0 and 140$^{\circ}$. Originally applied to RXTE data, the hue was also used for \nicer{} data by \citet{Wang_2022_revmachine} to determine and compare the accretion states of different sources. We follow their approach and calculate the hue with the 4.8-9.6 keV energy band. The interpretation of the hue is similar to that of the HR, but the hue is sensitive to changes in the timing properties, which may evolve more than the spectral shape during parts of the outburst. This is also illustrated in Table \ref{tab:obse_overview}, where the HR changes very little, but the hue covers a broad range of values. Both the HR and the hue are associated with changes in coronal geometry and we expect the spectral-timing properties to be consistent for observations with similar values of those properties.

We use 256 s segments to measure the hue of each observation (following \citet{Heil_2015} and \citet{Wang_2022_revmachine}) and obtain spectral-timing properties down to 0.0039 Hz. We will refer to the observations of \source{} by the last three digits of their observation IDs, which can be obtained by adding 1200120 before those three digits. We define five energy bands, which are used throughout this work, listed in Table \ref{tab:band_names}. 

The full list of observations used for the three sources can be found in Table \ref{tab:obse_overview}. In the table, we also show the QPO frequency, power-spectral hue, 4-10 / 2-4 keV HR and \nicer{} count rates normalised to 52 functioning FPMs.

\section{Analysis and results}
\label{sec:analysis_results}

\subsection{Flux-dependence of the QPO lags}
\label{subsec:fluxdep_lags}

\begin{figure*}
    \centering
    \includegraphics[width=176mm]{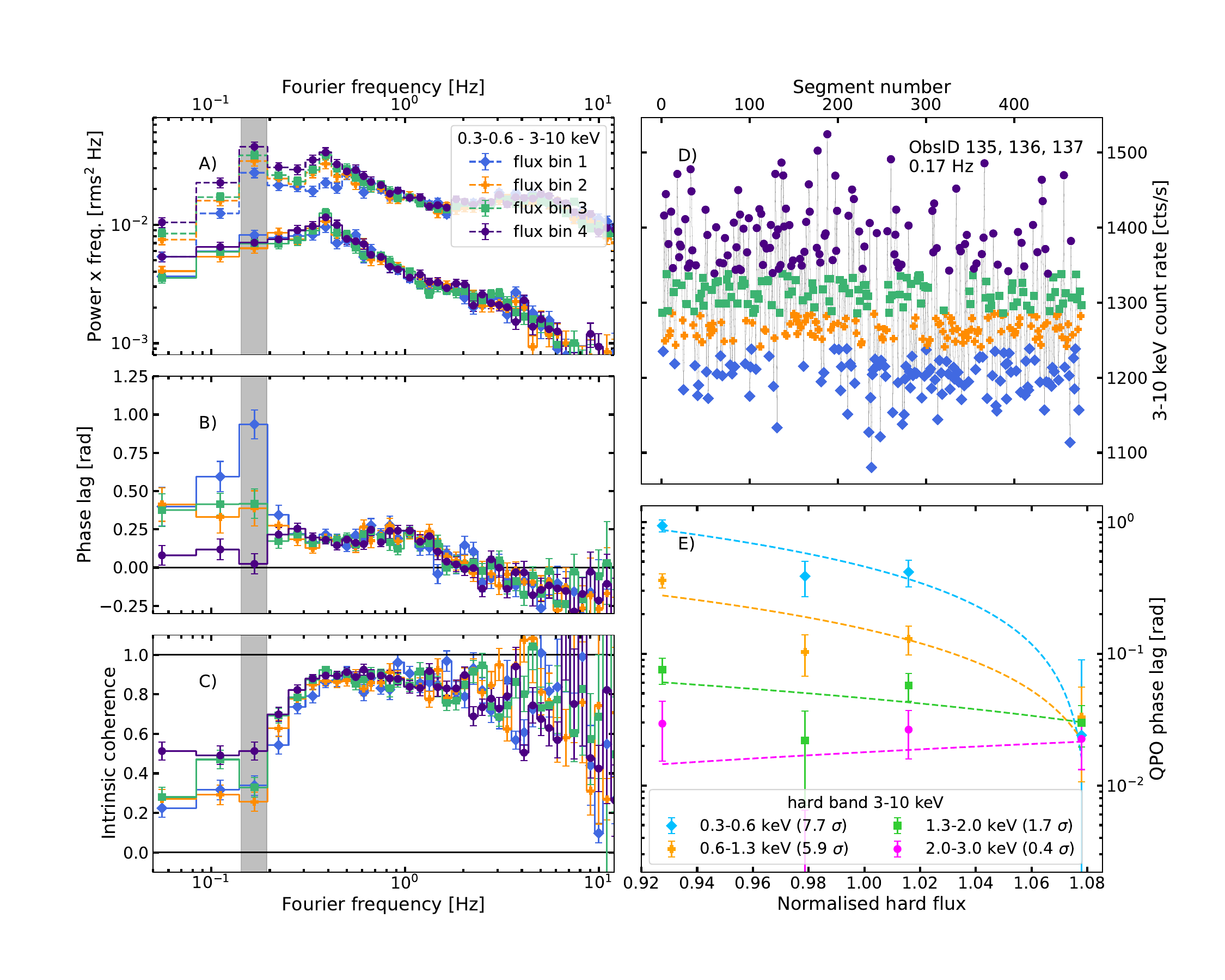}
    \caption{The five panels shown in the figure summarise the methods and results for finding the relation between the 3-10 keV flux and the QPO lags in observations 135, 136 and 137, which all have a QPO fundamental frequency of $\sim$0.17 Hz, indicated with the grey shaded area. In the left column, we show the power-spectra, phase lag- and coherence versus frequency spectra for four hard flux bins, using the very soft 0.3-0.6 and 3-10 keV energy bands. The power spectra for the hard band are shown with dashed lines, while the very soft band is shown with solid lines. The lags at the QPO frequency depend strongly on the instantaneous hard flux. The QPO lags for the very soft and hard band are largest when the hard flux is low and vice versa.
    The upper right panel shows the compacted light curve of the hard band, divided into four flux bins, each colour-coded in the same way as the spectral-timing properties on the left. The lower right panel shows the QPO lag - flux relation for different soft(er) bands. The relation is stronger for softer bands, while the lags between the 2-3 and 3-10 keV bands, both dominated by power-law emission, do not change significantly.}
    
    \label{fig:5panel}
\end{figure*}

\begin{figure}
    \centering
    \includegraphics[width=\columnwidth]{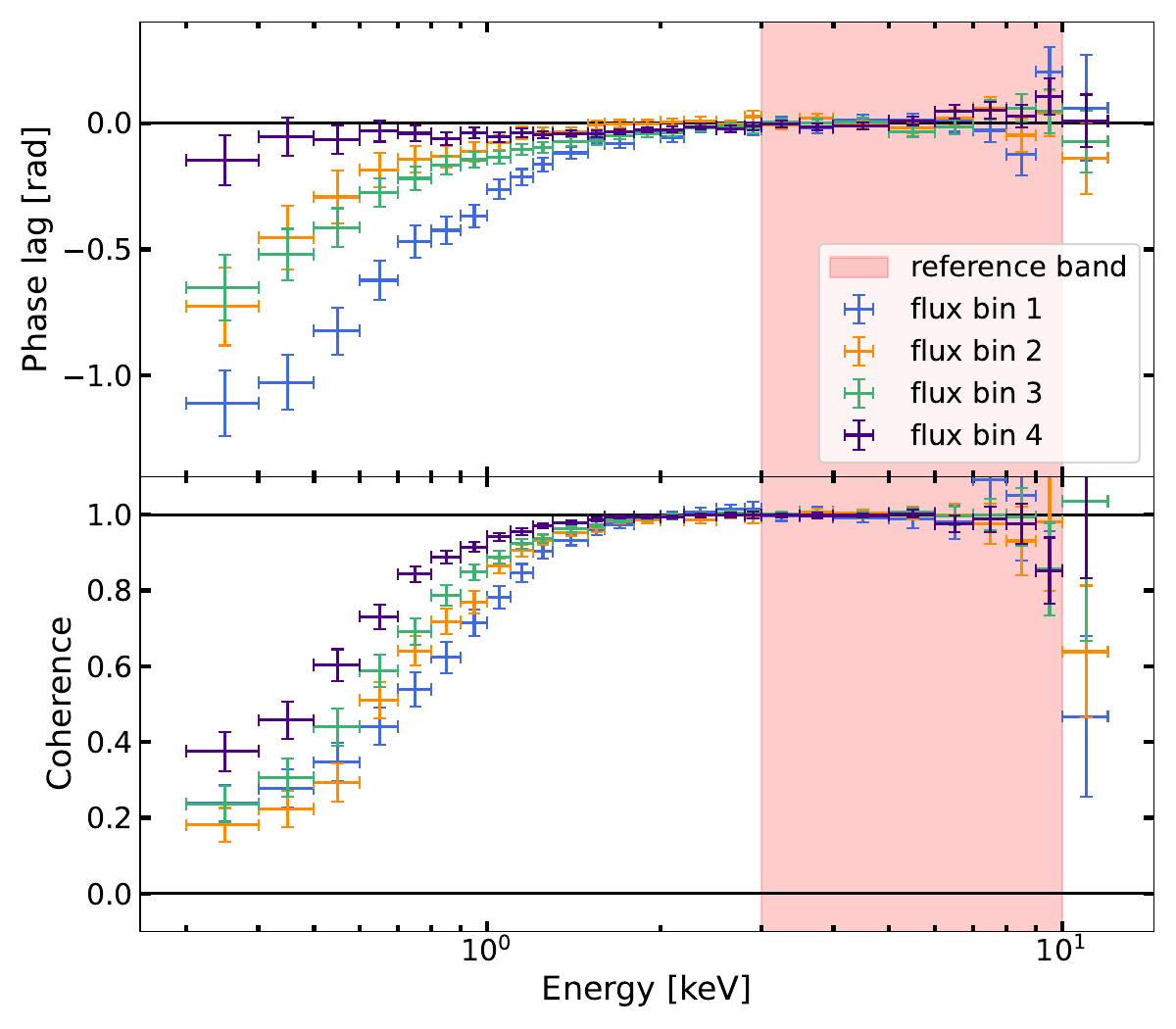}
    \caption{QPO lag- and coherence energy spectra for ObsIDs 135, 136 and 137 are shown for four hard flux bins in the top and bottom panels. The hard band is always 3-10 keV. It is clear that the lags start to deviate with flux below 1.3 keV, where disc emission is important. We follow the convention that hard lags move in the positive direction (upwards) with increasing energy, so the negative values at energies below 3 keV represent hard lags here.}
    \label{fig:lagcoh_energy}
\end{figure}

\begin{figure*}
    \centering
    \includegraphics[width=176mm]{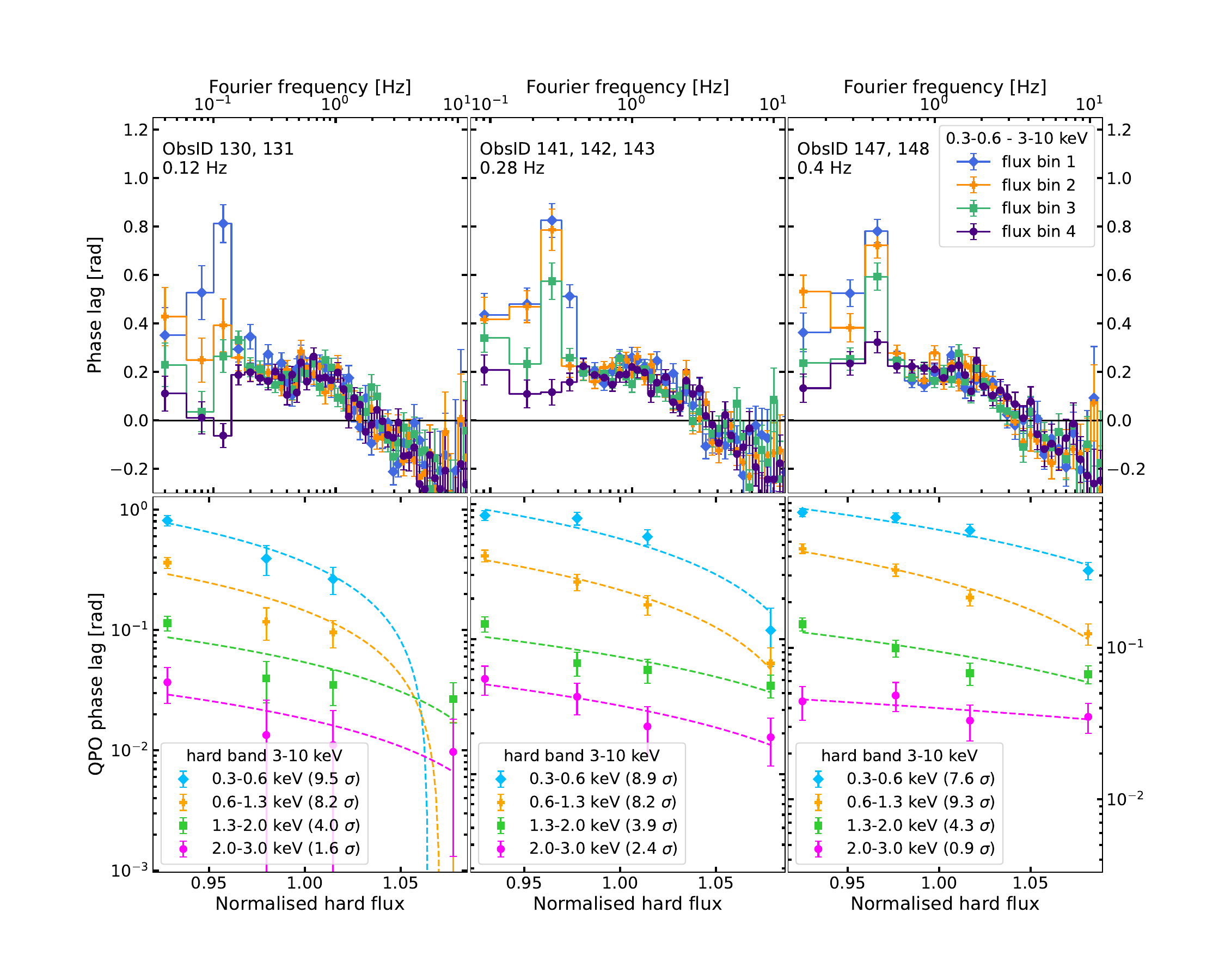}
    \caption{The upper panels show lag-frequency spectra between the very soft and hard band for three groups of observations with different QPO frequencies in four hard (3-10 keV) flux bins. It is clear that the lags at (and below) the QPO frequency depend strongly on the hard flux, with a clear peak of almost 1 rad at the QPO frequency for the lowest flux bin, while there is no feature for the highest flux bin. The lower panels quantify the relation between the QPO lag and the flux for four different soft bands. There is a strong relation between lag and flux for the soft bands, while lags between the medium bands and hard band have a much weaker link. We note that the values of $\sigma$ can indicate a significant relation between the QPO lag and hard flux, but their exact value should be interpreted with caution as the assumption of normally distributed errors breaks down far from the mean. For parameter values of the linear model, we refer to Table \ref{tab:slopes}.}
    \label{fig:6panel}
\end{figure*}

\begin{figure}
    \centering
    \includegraphics[width=\linewidth]{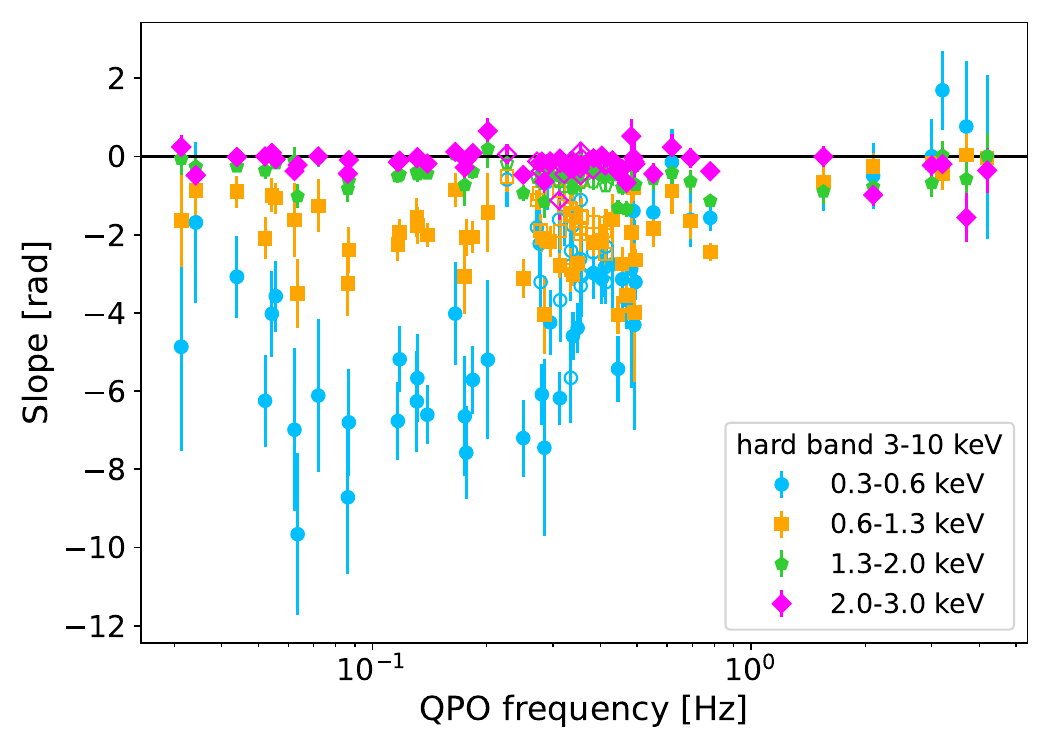}
    \caption{The relation between the slope of the QPO lag -- flux relation versus the QPO frequency for all observations in the hard state and HIMS of \source{} (ObsIDs 106-196). Almost all slopes are negative (harder lags for low hard flux) and for the very soft band the relation is much steeper below 0.3 Hz. The empty symbols correspond to the `bright decline' phase of the outburst, where the QPOs are weaker. }
    \label{fig:slopes_QPOf}
\end{figure}

\begin{figure*}
    \centering
    \includegraphics[width=176mm]{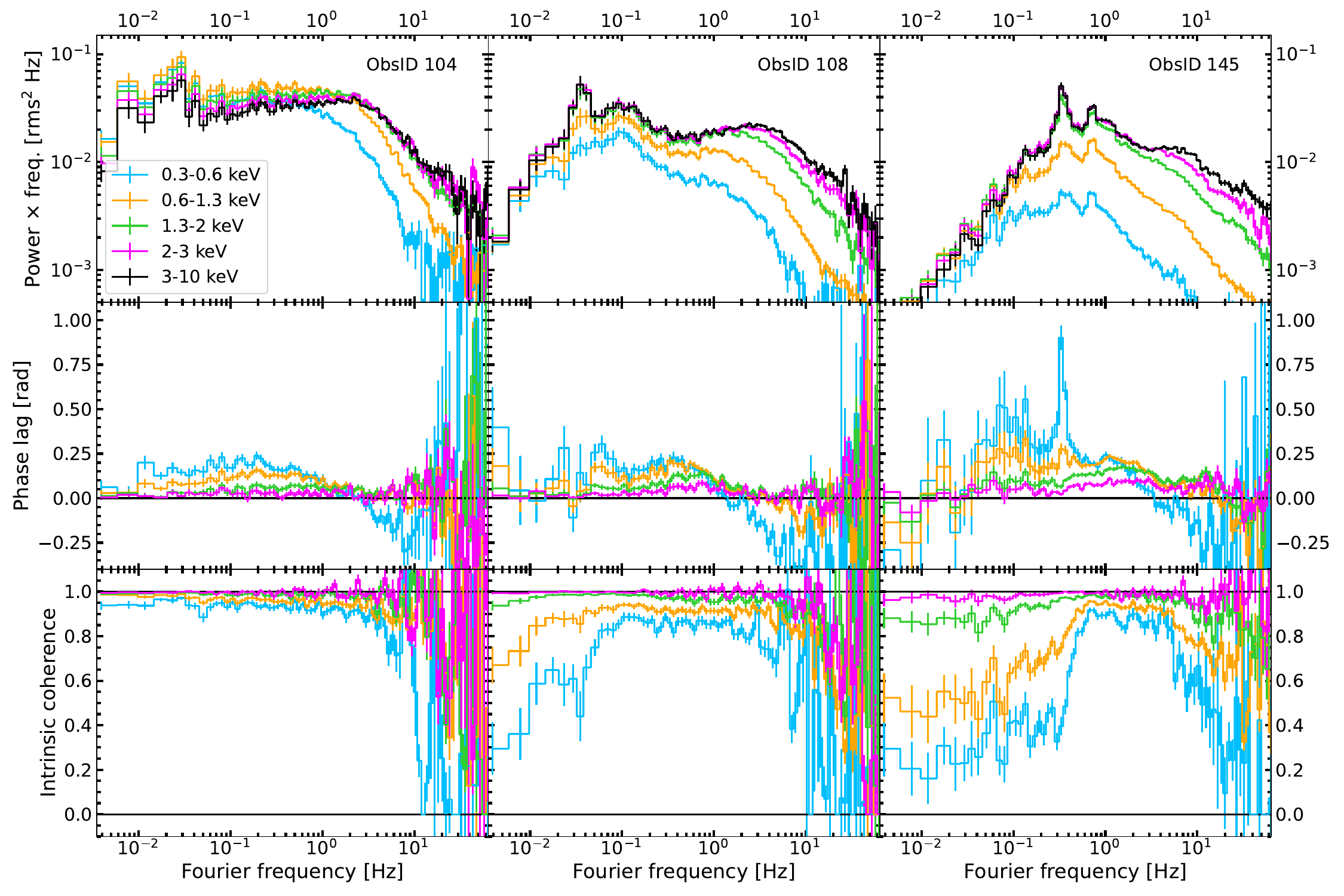}
    \caption{Power spectra, phase lag and coherence versus frequency spectra for different combinations of energy bands of observations 104 (left), 108 (middle) and 145 (right). The lags and coherence all have 3-10 keV as their hard band, while the different colours are defined by the softer band. In observation 104 (left column), there is no QPO and the coherence is high for all combinations of energy bands. The middle column shows observation 108, taken a few days after 104. There is a type-C QPO centered at $\sim0.034$ Hz, which is accompanied by a drop in the coherence at the QPO frequency and below, when comparing disc bands to a 3-10 keV hard band. On the right, the power spectra of observation 145 show a clear type-C QPO at $\sim0.34$ Hz, which is accompanied by a drop in the coherence at the QPO frequency and below for the softest bands. A clear peak at the QPO frequency is also visible for the 0.3-0.6 keV versus 3-10 keV phase lag.}
    \label{fig:9panel}
\end{figure*}

\begin{table}
    \centering
    \begin{tabular}{c|c|c|c|c}
       \textbf{ObsID} &\textbf{130, 131} & \textbf{135-137} &\textbf{141-143}& \textbf{147, 148}\\
       \hline
      Softer band  & Slope [rad] &Slope [rad]&Slope [rad]&Slope [rad]\\
      \hline
       0.3-0.6 keV & -5.7$\pm$0.6 & -5.7$\pm$0.7  &-5.1$\pm$0.6& -3.0$\pm$0.4
\\
       0.6-1.3 keV & -2.05$\pm$0.25 & -1.69$\pm$0.29 &-2.19$\pm$0.27&-2.03$\pm$0.22
 \\
       1.3-2 keV &-0.46$\pm$0.12&-0.2$\pm$0.12&-0.43$\pm$0.11&-0.43$\pm$0.10
 \\
       2-3 keV   & -0.15$\pm$0.09& 0.05$\pm$0.10 &-0.20$\pm$0.09&-0.08$\pm$0.08 \\
          
    \end{tabular}
    \caption{The slopes of the linear fit to the QPO lag -- flux relation as shown in Figs. \ref{fig:5panel} and \ref{fig:6panel} for all four groups of observations.  The hard band used is always 3-10 keV and the unit of the slope parameter is rad / normalised flux. The 1-$\sigma$ errors on the slope parameter were determined with a grid search.}
    \label{tab:slopes}
\end{table}

We extend the research presented in \citet{Bollemeijer_2024}, where we found that the short-term time lags in \source{} depend on the instantaneous hard (3-10 keV) X-ray flux. The lags in the 4-20 Hz range are associated with reverberation of coronal photons on the disc \citep{Kara_2019,Uttley_2025}, which depends strongly on the coronal geometry. As such, we interpret the variations in the short-term lags as being due to changes in the coronal geometry. The lags were found to vary both on the QPO time-scale, consistent with a geometric origin of QPOs, and on other time-scales. Here, we investigate whether the QPO lags themselves also vary on time-scales slightly longer than the QPO time-scale. Because QPO lags are thought to depend on coronal geometry as well, any relation between flux and QPO lags may indicate again that the geometry of the corona varies on a wide range of time-scales.

We follow a similar method to \citet{Bollemeijer_2024}, consisting of the following steps. First, we created light curves for all observations for the full energy band (0.5-10 keV). We measured the QPO frequencies reported in Table \ref{tab:obse_overview} by fitting two narrow Lorentzian functions to the fundamental and harmonic peaks of the QPO, and three broad Lorentzians to the broadband noise. We combined observations with similar QPO frequencies to increase the signal of our measurements, which are also listed in Table \ref{tab:obse_overview}. For each observation, we measured the number of FPMs that were switched on and passed the screening with \texttt{nicerl2}. We then created light curves for the energy bands listed in Table \ref{tab:band_names} and normalised them by the number of FPMs that were used in each observation.

To study the flux-dependence of the QPO lags in \source, we made shorter light curve segments of three times an average QPO cycle. For the combined observations shown in Fig. \ref{fig:5panel}, 135, 136 and 137, which have a QPO frequency of $\sim0.17$ Hz, this corresponds to a segment length of 18 s. We grouped these 18 s light curve segments into four hard (3-10 keV) flux bins, which is illustrated in panel D of Fig. \ref{fig:5panel}, where the different colours and markers indicate the different flux bins and each data point represents a 18 s segment. When comparing the hard fluxes within an observation, it is important that there is no strong long term change in count rate, which would dominate over the variability on the order of the segment length of 18 s. From panel D in Fig. \ref{fig:5panel}, it is clear that there is no dominant long term trend and we can use the method. An important difference between the methods presented in \citet{Bollemeijer_2024} and the current work, is that in the previous paper we compared the flux in 0.25 s slices \textit{within} 64 s segments to create flux bins, while in the current work we compare the flux in 18 s segments within an observation or even multiple observations, as is illustrated in panel D of Fig. \ref{fig:5panel}. As such, we probe rather different time-scales than in \citet{Bollemeijer_2024}.

After flux binning, we have four sets of light curve segments in different energy bands, for which we calculate several spectral-timing properties. We calculated the frequency-dependent power spectra, phase lags and coherence following \citet{Uttley_2014review} for all four flux bins. For the cross-spectral properties, the phase lags and the coherence, we used different soft bands (see Table \ref{tab:band_names}) and a 3-10 keV hard band. In panel A, B and C, in the left column of Fig. \ref{fig:5panel}, we show the power spectra, phase lag and coherence versus frequency spectra for the very soft and hard bands for the different hard flux bins, allowing us to study how these properties depend on the hard X-ray flux. In all lag versus frequency plots, positive lags denote hard lags, as is convention.

We find that there is some flux-dependence of the hard band fractional rms at low frequencies, as is visible in panel A of Fig. \ref{fig:5panel}, while the very soft band power spectra show little change. The hard power spectral change may be related to the rms-flux relation \citep{Uttley_2001,Uttley_2005,Heil_2012}, which leads to a larger absolute rms for higher fluxes. The flux-dependence of the fractional rms (which we plotted in panel A of Fig. \ref{fig:5panel}) is less straightforward to predict from the rms-flux relation and depends on source state, but \citet{Heil_2012} found that in the hard state, higher flux will also lead to larger fractional rms, which is also what we observe. We note that the QPO rms follows a more complicated relation with flux \citep{Heil_2011}. In the case of \source, the relation between the QPO fractional rms and the flux could be similar to the surrounding broadband noise, which is consistent with the lowest frequency QPOs in BHXRB XTE~J1550-564 as investigated by \citet{Heil_2011}.

Panel B shows the QPO lags for different flux bins and we see a much more dramatic effect. We define QPO lags here as the phase lags in the third Fourier frequency bin when using light curve segments of $\sim$ three QPO cycles, as that frequency bin is dominated by the QPO signal. At low hard flux, there is a strong QPO lag feature, reaching almost 1 rad in amplitude, while the QPO lag is consistent with zero for the high flux bin. As such, the difference in lags between the lowest and highest flux bin is almost 1 rad. Below the QPO frequency, the lags also seem to depend on the hard flux, although the change in lag amplitude is smaller.
The frequency range where the lags depend on the hard flux coincides with a steep drop in the intrinsic coherence (see panel C of Fig. \ref{fig:5panel}) at and below the QPO frequency. The coherence itself has a weak dependence on the flux, with the highest hard flux bin showing a slightly higher coherence than the other bins.

To quantify the relation between the QPO lags and the hard flux, we fit a simple linear model to the QPO lag versus hard flux, as shown in panel E of Fig. \ref{fig:5panel}. The panel shows the phase lags at the QPO frequency for different soft bands versus the hard 3-10 keV band on the y-axis and the normalised 3-10 keV count rate on the x-axis. The y-axis is scaled logarithmically to show the slopes in all four soft energy bands, which is difficult to see with a linearly scaled axis. By fitting a constant and a linear model to the lag versus flux data and calculating the difference in $\chi^2$, we estimate the significance of the relation between the quantities. If the constant model describes the data well, the $\Delta\chi^2$ is drawn from a $\chi^2_1$-distribution with one degree of freedom, as the linear model has one more free parameter. We assume here that the errors on the phase lags are normally distributed, which is reasonable as we calculate the lags from averaging over the cross-spectrum of at least 100 segments per flux bin \citep{Huppenkothen_2018,Ingram_2019formulae}. The values of $\sigma$ reported in panel E of Fig. \ref{fig:5panel} represent the probability of obtaining the data if there were no relation between the QPO lag and the hard flux. For the softest bands, there is a significant relation between both ($>5\sigma$), while the medium bands (1.3-3 keV) show slopes that are less steep and $<3\sigma$ significant. The reported values of $\sigma$ can indicate that there is a significant relation between the QPO and the flux, but their absolute value should be interpreted with caution, as the assumption of normally distributed errors on the lags breaks down beyond a few $\sigma$. We repeated the analysis above using longer segment lengths (e.g. six QPO cycles), which returned similar results. However, the smaller flux range and number of usable segments led to larger error bars, so we only present results using segment lengths of approximately three QPO cycles.

The energy-dependence of the QPO lag -- flux relation becomes clearer in a lag versus energy spectrum \citep{Uttley_2014review}. In Fig. \ref{fig:lagcoh_energy}, we show QPO lag- and coherence versus energy spectra for four flux bins, with a hard 3-10 keV reference band, for ObsIDs 135, 136 and 137 combined. Again, we use 18 s segments and show the results obtained by averaging over the cross-spectrum (and the power spectra when calculating the coherence) of the third Fourier-frequency bin, which is dominated by the QPO signal. For the errors on the lags and coherence, we follow \citet{Ingram_2019formulae} and \citet{Vaughan_1997coherence}, respectively. In the top panel of Fig. \ref{fig:lagcoh_energy}, the lags in different flux bins start to deviate below $\sim$1.3 keV, where disc emission becomes important. In the figure, we use the convention that positive values indicate that the photons arrive later in a given energy band, so the negative values at energies below the reference band indicate hard lags (soft photons arrive earlier than hard photons). The highest flux bin has QPO lags consistent with 0 rad, while the lowest flux bin shows hard lags with an amplitude of $>$1 rad. The coherence decreases below $\sim$1.3 keV for all four flux bins, although flux bin 4 (high hard flux) has a slightly higher coherence than the other bins.

To show that the QPO lag -- flux relation is observed at different QPO frequencies in \source, we show the lag-frequency and QPO lag -- flux (panels B and E of Fig. \ref{fig:5panel}) for different sets of observations in Fig. \ref{fig:6panel}. At QPO frequencies of 0.12, 0.28 and 0.4 Hz, we measured significant anticorrelations between the hard flux and the QPO lags for the soft bands, as is also visible in the slope parameter values of the linear fit for all four groups of observations presented in Table \ref{tab:slopes}. We determined the 1-$\sigma$ errors on the slope parameter with a grid search. In most cases, the medium-soft band (1.3-2 keV) seems to show a significant relation, although with less certainty than the softer bands. In all analysed observations, the lags below the QPO frequency show a trend that is similar to what we observe at the QPO frequency, with the lowest hard flux bin showing the largest lag amplitude. The energy-dependence of the QPO lag -- flux relation may indicate that in particular the disc--power-law lags, as opposed to the power-law--power-law lags, are anticorrelated with the hard flux, potentially constraining the mechanisms causing such changes in the lags, and we will discuss a similar interpretation in Section \ref{sec:discussion}. 

We also tested whether the observed QPO lag --flux relation depends on QPO frequency. To investigate such a dependence, we followed the methods outlined earlier on all \nicer{} observations of \source{} in the hard state (observations 106-196). After finding the QPO frequency for each observation by fitting Lorentzians, we determined the slope between the QPO lag and the flux for the four softer bands and the hard 3-10 keV band, using segments of $\sim3$ QPO cycles. The result is shown in Fig. \ref{fig:slopes_QPOf}, where we can see steep negative slopes (harder lags for lower hard flux), especially below $\sim$0.3 Hz. Above 0.5 Hz, the relation is much weaker. In the figure, the empty symbols between 0.2 and 0.4 Hz correspond to the `bright decline' phase of the outburst, as defined by \citet{De_Marco_2021}, where the luminosity drops by a factor of a few, while the spectral hardness does not change much. During the bright decline, the QPOs are much weaker, so it is perhaps not surprising that the measured QPO lag -- flux slope is less steep during this part of the outburst.
% From Table \ref{tab:slopes}, no obvious trend with QPO frequency can be discerned, although the included observations span a limited range in QPO frequency of 0.12 to 0.4 Hz.

\begin{figure*}
    
    \centering
    \includegraphics[width=176mm]{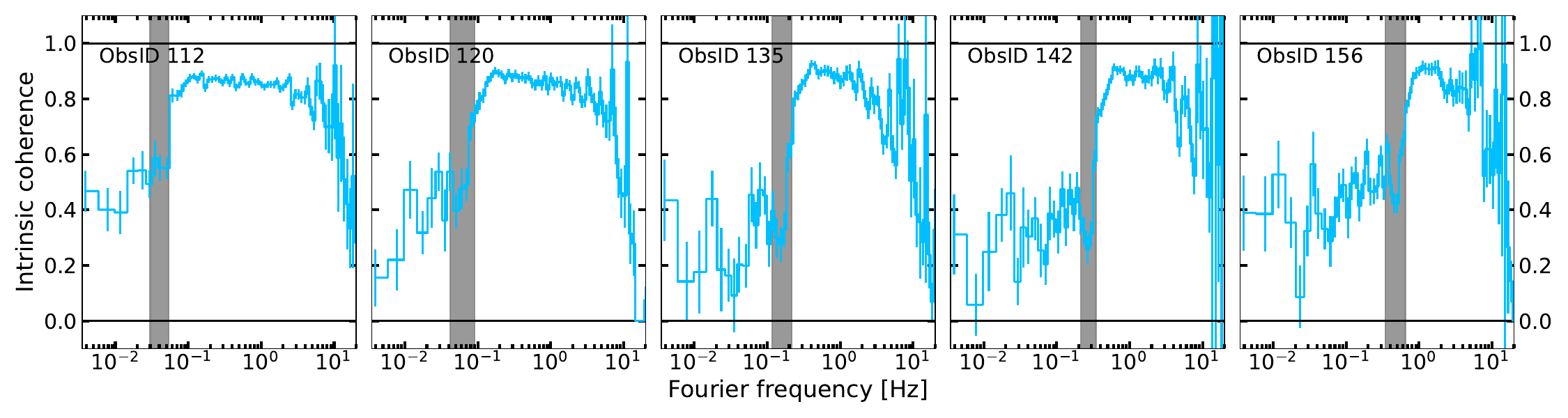}
    \caption{The coherence versus frequency spectra for the very soft (0.3-0.6 keV) and hard (3-10 keV) band for five observations with increasing QPO frequency. The grey shaded areas show the QPO centroid frequency $\pm$ HWHM. It is clear the QPO frequency marks the border of the low and high coherence frequency ranges and the coherence pattern closely follows the QPO frequency as it increases during the outburst.}
    \label{fig:5cohpanel}
\end{figure*}

\subsection{Energy-dependence of the lags and coherence}
\label{subsec:engdep_lags_coh}

To investigate the relation with the QPO and the energy-dependence of the phase lags and coherence further, we show the average power spectra, phase lag- and coherence versus frequency spectra of three observations in Fig. \ref{fig:9panel}. The first observation is 104, in the left column of Fig. \ref{fig:9panel}, which was made on 15 March 2018 (MJD 58193), and shows \source{} in the rising hard state. No QPOs are visible in the power spectrum for any energy band and the power spectra are similar for all energies, although the variability in the softest band is clearly suppressed at high frequencies, which is expected for emission from an accretion disc \citep{Uttley_2025}. The hard lags at frequencies below $\sim2$ Hz increase smoothly with decreasing energy and for the very soft band, we see soft lags at high frequencies. The coherence is high, almost unity, for all energy bands.

The picture is very different in the middle column of Fig. \ref{fig:9panel}, which shows observation 108, taken a few days later on 23 March 2018 (MJD 58200). A weak type-C QPO is visible at $\sim0.034$ Hz. Although the lags at broadband noise frequencies look qualitatively similar to those measured in observation 104, the lags at the QPO frequency are close to zero for all energy bands. More significantly, the coherence is decreased for the softest energy bands versus the 3-10 keV band at and below the QPO frequency. The relation to the QPO is clearer in the right column of Fig. \ref{fig:9panel}, for observation 145, where a larger decrease in coherence is visible below the QPO frequency of 0.34 Hz. In all cases, the coherence between the medium and hard bands (above $\sim$1.3 keV) is close to unity, while the soft bands show more complex relations with the hard emission. In the right column of Fig. \ref{fig:9panel} the 0.3-0.6 versus 3-10 keV phase lag shows a clear peak at the QPO frequency, which is not visible in the other energy bands. 

The intrinsic coherence between the soft bands and the hard 3-10 keV band is far from unity at and below the QPO frequency. To demonstrate how the coherence and the QPO are related, we show in Fig. \ref{fig:5cohpanel} how the drop in very soft versus hard band coherence clearly follows the QPO frequency as it increases during the outburst. The QPO centroid frequency $\pm$ the half width at half maximum (HWHM) from a multi-Lorentzian fit is shown as a grey shaded area for the five \source{} observations. The coherence below and at the QPO frequency is $\lesssim0.5$ and increases to values above 0.8 at slightly higher frequencies. 
To our knowledge, such a link between the low-frequency coherence and the QPO frequency has not been observed before. 
We suggest several possible interpretations of the new phenomenon and its potential link to the  QPO lag -- flux relation in Section \ref{sec:discussion}.

\subsection{Comparison with \sourcem{} and \sourceg}
\label{subsec:1803_lagcoh}

\begin{figure}
    \centering
    \includegraphics[width=\linewidth]{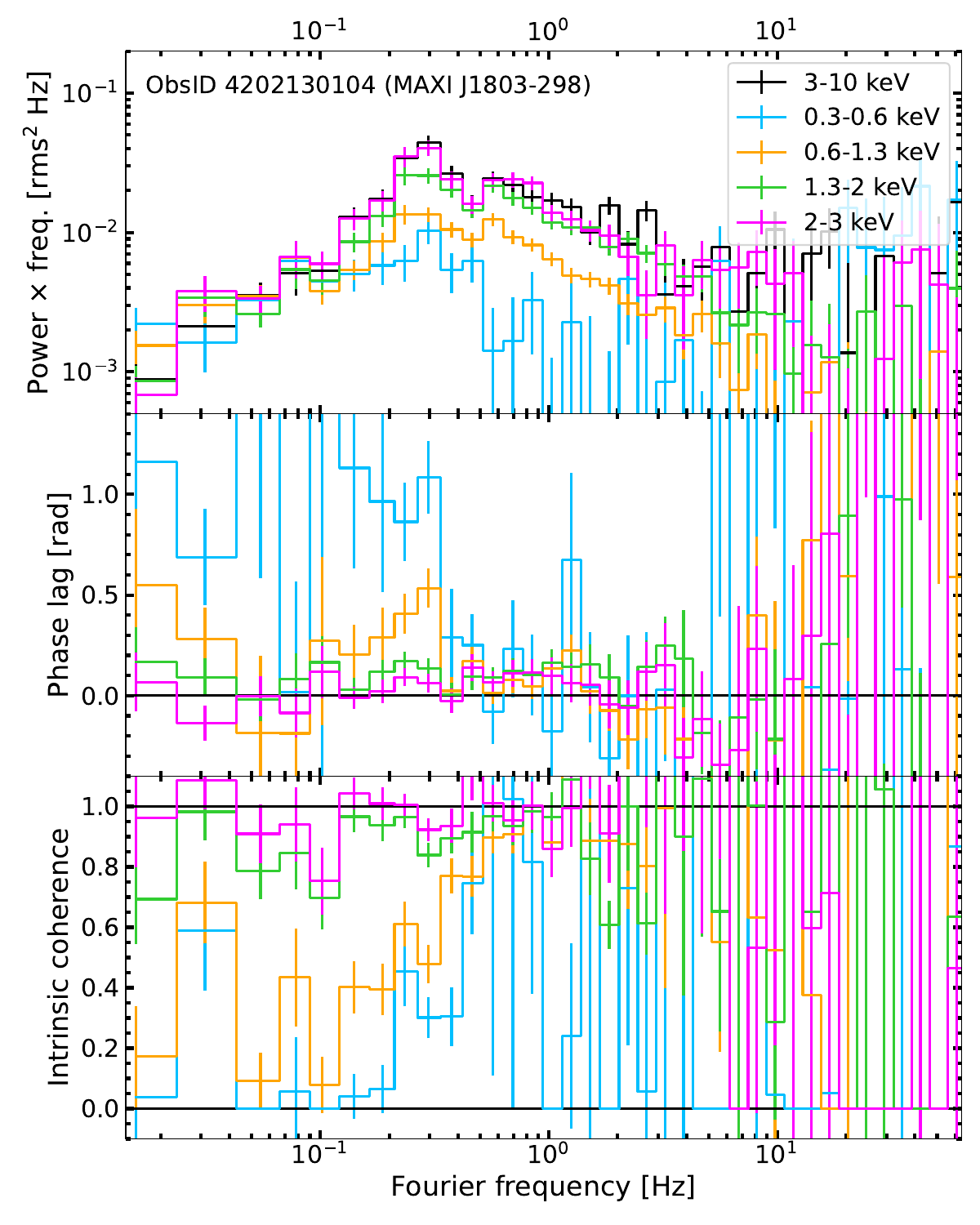}
    \caption{The power spectra and phase lag and coherence versus frequency spectra for a subset of data from observation 4202130104 of \sourcem. At the QPO frequency, around 0.27 Hz, a clear increase in the phase lags between the soft bands below 1.3 and the 3-10 keV hard band is visible. Also, the coherence at and below the QPO frequency is low for those combinations of energies, while the coherence is close to unity above the QPO frequency, very similar to \source{} in the right column of Fig. \ref{fig:9panel}. When comparing the figures, note that the segment size used for \sourcem{} is 64 s, as opposed to the 256 s used for \source. We only show the results for the first 15 segments here, because the QPO frequency exhibits significant drift during this observation. The large error bars for the very soft band are due to the higher interstellar absorption in \sourcem.}
    \label{fig:pslagcoh_1803}
\end{figure}

\begin{figure}
    \centering
    \includegraphics[width=\linewidth]{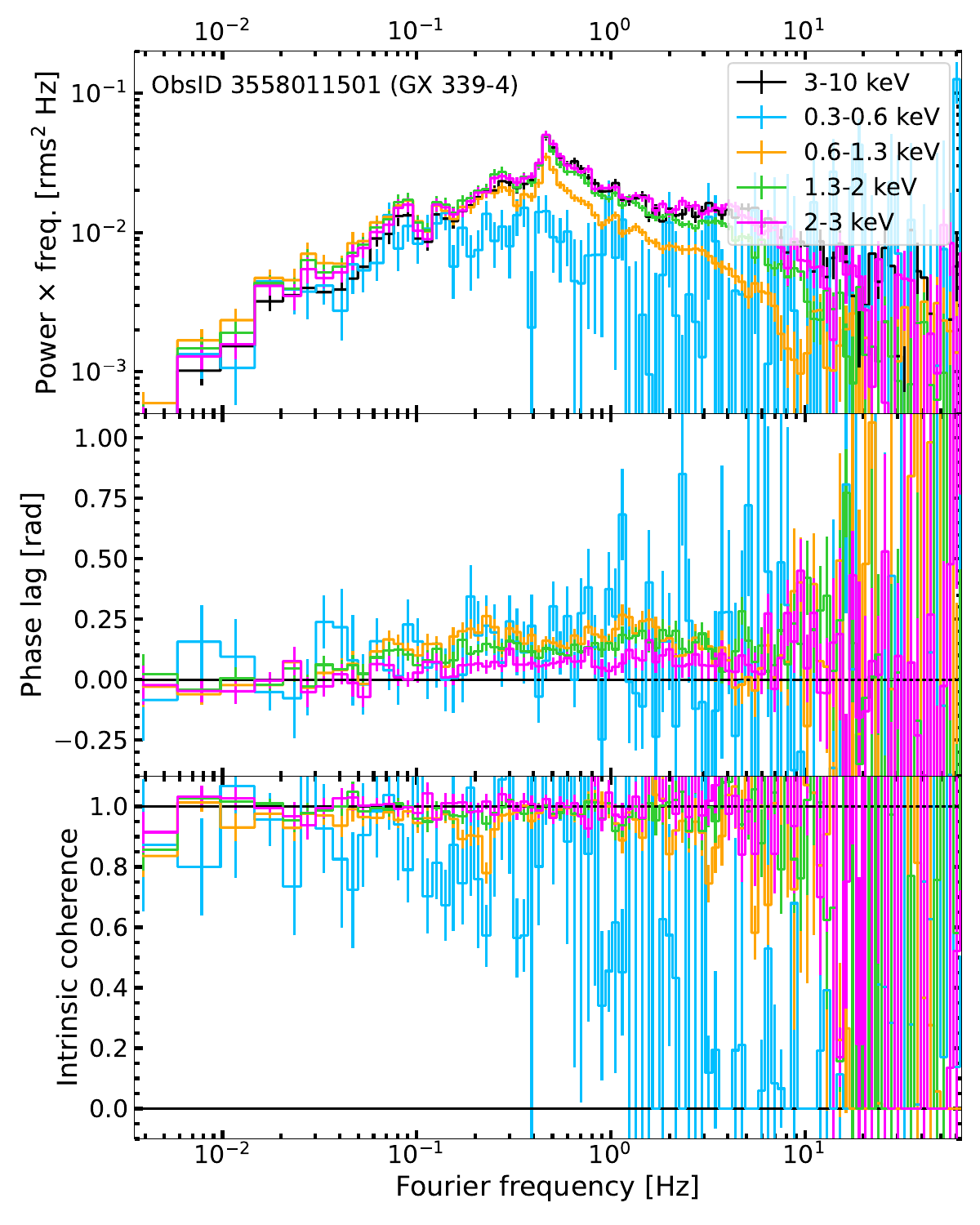}
    \caption{The power spectra and phase lag and coherence versus frequency spectra from observation 3558011501 of \sourceg. The coherence is high for all energy bands and no lag feature can be seen at the QPO frequency of 0.45 Hz. The very soft band timing properties have large error bars due to the higher interstellar absorption in \sourceg. }
    \label{fig:pslagcoh_gx339}
\end{figure}

In \source, we find a relation between the hard flux and the QPO lags, especially for (very) soft and hard bands (as opposed to medium and hard bands). Also, we observe a strong drop in coherence between soft and hard energy bands at and below the QPO frequency. In \citet{Cassatella_2012}, a drop is seen in the low-frequency coherence for the softest energy band using XMM-Newton data of the hard state of Swift~J1753.5-0127, but there is no clear link with a QPO in their analysis. To test whether the observed behaviour is unique to \source, we investigated \nicer{} data for two other BHXRBs: \sourcem{} and \sourceg, in similar accretion states. We first compare observations of \sourcem{} and \sourceg{} with observation 145 of \source, which has an HR of 0.2906 $\pm$ 0.0002 and a power-spectral hue of $129.1 \pm 1.0^\circ$ (see Table \ref{tab:obse_overview}).

We analysed \nicer{} observations of BHXRB \sourcem{}, which has a high inclination of around $70^\circ$ \citep{Adegoke_2024}. The interstellar absorption ($\rm{N_H} \sim 3.2\times10^{21} \rm{cm}^{-2}$, \citealt{Adegoke_2024}) is not quite as low as for \source, but still allows reliable measurements with soft bands. \sourcem{} shows periodic dips in the X-ray light curve as the bulge, where the accretion stream from the companion star impacts the disc, blocks the X-ray emitting region near the black hole from our view \citep{Jana_2022}. We excluded those dips from our analysis of \nicer{} observation 4202130104, made on May 5, 2021 and shown in Fig. \ref{fig:pslagcoh_1803}. The HR for this observation is 0.334 $\pm$ 0.0008 and the power-spectral hue $110\pm5^{\circ}$. The three panels show the power spectra, phase lag and coherence-frequency spectra for different energy bands, using 64 s segments (as opposed to the 256 s segments for \source). Because the QPO frequency evolves during the observation, the QPO peak is very broad when using all data, so we only show the spectral-timing properties for the first 15 segments of observation 4202130104, with a QPO frequency of 0.27 Hz. The very soft 0.3-0.6 keV energy band is more absorbed than for \source{} and the count rate is about an order of magnitude lower, which explains the large error bars for those energies. Following Fig. \ref{fig:9panel}, the phase lag and coherence are calculated using the same 3-10 keV hard band and using different soft bands. Comparing Fig. \ref{fig:pslagcoh_1803} to the middle and right column of Fig. \ref{fig:9panel}, it is clear that two properties of \source{} can also be observed in \sourcem. First, the phase lags for soft (<1.3 keV) bands and the hard band show a clear peak at the QPO fundamental frequency of $\sim$0.27 Hz, while such a feature is absent in the lags between the medium (1.3 keV < E < 3 keV) and hard bands. Also, the coherence for the soft bands is low at and below the QPO frequency, while harder energy bands remain highly coherent. As the QPO frequency increased during the start of the outburst, the coherence pattern follows in the same way as observed in \source, which illustrated in Appendix \ref{app_J1803}, where we show the spectral-timing properties of observation 4202130102 of \sourcem{}, with a lower QPO frequency of 0.13 Hz. The low-frequency coherence seems to be even lower in \sourcem{} than in \source. The lower count rates and higher absorption impact our ability to measure the coherence, especially for the soft band, but we see the same phenomenon in the soft (0.6-1.3 keV) band. 

We also attempted to measure a QPO lag versus hard flux relation in \sourcem, and in a few cases obtained a marginally significant (3$\sigma$) preference for a linear over a constant model fit. The measured slopes for the respective soft bands are -1.1 $\pm$ 1.9, -2.8 $\pm$ 0.9, -1.2 $\pm$ 0.5 and -1.1 $\pm$ 0.4 rad for all (non-dipping) segments in observation 4202130104 of \sourcem, using an average QPO frequency of 0.32 Hz for this observation. The improvement of a linear model compared to a constant model is 3.3$\sigma$ significant for the soft versus hard band, and less significant for the other bands. Because the QPO frequency drifts during observation 4202130104, we also calculated the slopes for the subset of data shown in Fig. \ref{fig:pslagcoh_1803}, with a QPO frequency of 0.27 Hz. The slopes are then -9.8 $\pm$ 3.3, -5.2 $\pm$ 1.9, -0.5 $\pm$ 1.0, 0.6 $\pm$ 0.9 rad, and the significance of the linear relation is just under 3$\sigma$ for the (very) soft bands. The slopes are generally negative and have similar values to those measured for \source, but the small amount of lower quality data prevents us from detecting the QPO lag -- flux relation with a high level of certainty in \sourcem. We require more observations and higher count rates to draw more definite conclusions, but given the lower data quality, the behaviour in \sourcem{} is consistent with what we measure in \source.

% $0.0 \pm 1.9$, $-1.0 \pm 0.8$, $-0.7 \pm 0.4$ and
% $-0.6 \pm 0.4$ wrong

For \sourceg, we analysed several \nicer{} observations during its outburst in 2021. In Fig. \ref{fig:pslagcoh_gx339}, we show our results for observation 3558011501, which was made on March 21, 2021. The HR is 0.3228 $\pm$ 0.0005 and the power-spectral hue is $90\pm3^{\circ}$ for this observation. \sourceg{} is a very well-studied source, as it goes into outburst every few years, and to our knowledge, no large drops in coherence related to the QPO were reported before. X-ray reflection fitting returns inclination values of 40-60$^{\circ}$, so \sourceg{} is considered a low-inclination source \citep{Garcia_2015,Zdziarski_2019}. The interstellar absorption ($\rm{N_H}\sim0.58\times10^{22} \rm{cm}^{-2}$, \citealt{Wang_2020}) is higher than in the other two sources considered in this paper, which is the main reason the error bars for the very soft band in Fig. \ref{fig:pslagcoh_gx339} are large. From Fig. \ref{fig:pslagcoh_gx339}, it is clear that there is no drop in coherence and no feature in the lags at the QPO frequency of 0.49 Hz. Despite the low count rates for the very soft band, any features comparable in size to those observed in \source{} and \sourcem{} should well be visible in the soft (0.6-1.3 keV) band. There is a hint of sub-harmonic structure at $\sim0.25$ Hz, which may be related to the small peak in the soft -- hard lags and the modest drop in coherence, but the effect is clearly much smaller than in the other two sources. We note that \citet{Buisson_2025} observed flip-flop behaviour in the type-B QPO of \sourceg, whose presence shows a strong dependence on source flux, providing new constraints on the mechanism causing type-B QPOs. However, we do not measure any significant dependence of the type-C QPO properties on flux in the hard state of \sourceg, as there is no significant change in the lags at the QPO frequency for different hard flux bins. Given the dependence of the QPO lag -- flux relation on QPO frequency in \source{} (see Fig. \ref{fig:slopes_QPOf}), we also investigated lower QPO frequency observations of \sourceg, where the relation may be more clear, but found that all observations showed spectral-timing properties very similar to those presented in Fig. \ref{fig:pslagcoh_gx339}.

\section{Discussion}
\label{sec:discussion}

In \source, we find that the QPO lag between disc and coronal power-law energy bands depends on the instantaneous power-law flux. When the hard flux is high, we see small amplitude hard lags, while the hard lags are much larger, up to 1 rad, when the hard flux is low, as was shown in Section \ref{subsec:fluxdep_lags}. At the same time, the coherence between disc and power-law bands is low at and below the QPO frequency in high-inclination sources \source{} and \sourcem, while it remains high in low-inclination source \sourceg. In this section, we will investigate several possible explanations for the observed behaviour.

\begin{figure}
    \centering
    \includegraphics[width=\columnwidth]{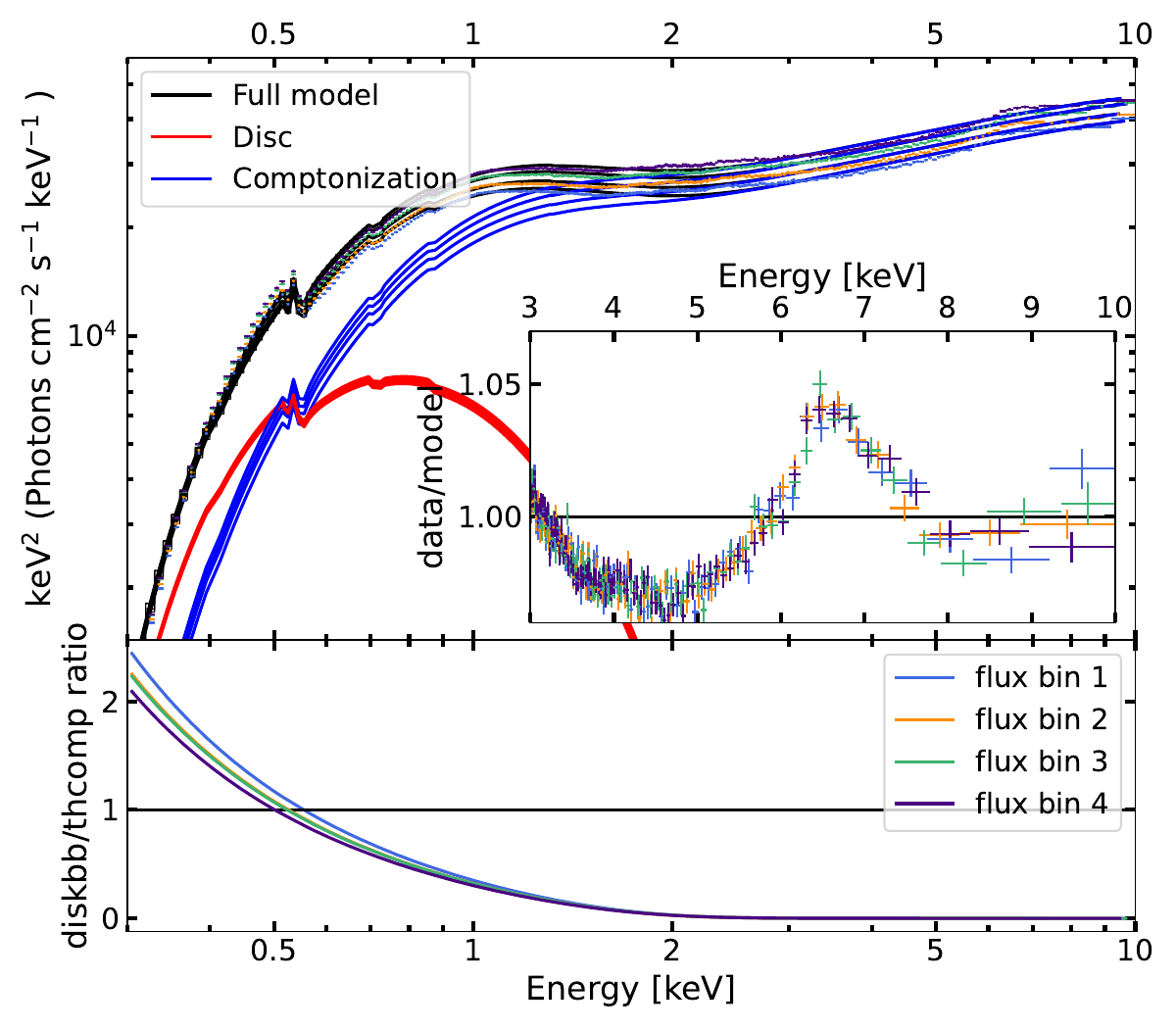}
    \caption{The upper panel shows the spectra of four hard flux bins of ObsIDs 135, 136 and 137, fitted with a \texttt{tbabs*(diskbb+thcomp*bbody)} model. The covering fraction of the \texttt{thcomp} component is set to 1, so all emission is from Comptonization, and the temperature of \texttt{bbody} is equal to \texttt{Tin} of the \texttt{diskbb}. The contribution from the \texttt{diskbb} is shown in red, while the Comptonization component is blue. The inset plot shows data/model ratio in the 3-10 keV, illustrating that the shape of the Fe K line does not change visibly. The lower panel shows the ratio of the fitted \texttt{diskbb} and \texttt{nthcomp*bbody} spectral components, which show only a small change ($\sim10\%$) at the softest energies.}
    \label{fig:fluxresolved_spectra}
\end{figure}

\subsection{Spectral changes due to flux binning}
\label{subsec:discussion_specchange}
First, we consider whether spectral changes from the grouping of light curve segments by their hard flux could explain the observed changes in the QPO lags. If the intrinsic lags between different spectral components (e.g. disc and power-law) are constant in time, but the relative contribution of those components to different energy bands varies (as we are binning on flux), the measured lags could vary as well. On the other hand, the intrinsic lags themselves may also be changing in time, but before we discuss scenarios for varying intrinsic lags, we study the possibility that spectral changes lead to variations in the measured lags. 

The change in the lags at the QPO frequency is large, consistently showing a difference of $>0.5$ rad between the highest and lowest hard flux bin (see Figs. \ref{fig:5panel} and \ref{fig:6panel}). Since we combined different light curve segments based on their hard flux, the spectra of those segments will be affected and the very soft band is expected to contain a varying contribution of photons from the corona. In a simplistic scenario, where the disc versus power-law lags are much larger than the power-law versus power-law lags (e.g. see center right panel of Fig. \ref{fig:9panel}), variations in the contribution of the disc and power-law components to the soft band may cause changing lags (see also the discussion in \citealt{Bollemeijer_2024}). The direction of the QPO lag -- flux relation is consistent with such an explanation, showing larger lags (more disc versus power-law) for low hard flux bins.

To test whether spectral changes in the soft band can explain the variations in the QPO lags with hard flux, we analysed the spectra of four flux bins of ObsIDs 135, 136 and 137. We created spectra for all four flux bins and followed \citet{Koenig_2024} by adding a 5\% systematic error, which allows us to apply a simple model to parametrise the broad structure of the spectrum, ignoring reflection and narrower (calibration) features that we do not model. We fit all four flux binned spectra with the model \texttt{tbabs*(diskbb+thcomp*bbody)} \citep{Zdziarski_2020}. The \texttt{diskbb} component takes into account the emission from the disk, while we set the \texttt{cov\_frac} parameter of \texttt{thcomp} to 1, such that all emission from the \texttt{bbody} component is Comptonized. The temperature \texttt{kT} of the \texttt{bbody} component is tied to the parameter for the inner disk temperature \texttt{Tin} of \texttt{diskbb}, while the normalizations of \texttt{diskbb} and \texttt{bbody} are free to vary, as are \texttt{Gamma} and \texttt{Tin}. This simple spectral model mimics the situation where the corona receives most of its seed photons from the inner regions of the disk, where the spectrum is similar to a blackbody with temperature \texttt{Tin}.
The result of the four spectral fits is shown in Fig. \ref{fig:fluxresolved_spectra}, where the upper panel shows the spectra and the model fits. The relative contribution of the \texttt{diskbb} (red) and \texttt{thcomp} (blue) components changes on the order of $\sim10\%$ at the softest energies. The disc contribution to the very soft band is on the order of 50\%, while it is close to zero for the power-law-dominated medium-hard band (2-3 keV). 

Comparing the spectroscopic and timing properties of the four flux bins, we conclude the following. In the different flux bins, the contributions from the disc blackbody and Comptonized components are very similar, while the QPO lags change significantly. For example, the QPO lags of the very soft band in the highest hard flux bin (bin 4) are close to those in harder energy bands (see the lag versus flux panels in Figs. \ref{fig:5panel} and \ref{fig:6panel}), while for lower hard flux bins, the QPO lags are much larger. The spectra of the different flux bins, as shown in Fig. \ref{fig:fluxresolved_spectra}, show only small differences. We therefore conclude that the large variations in the lag amplitude cannot be explained by a simple scenario where the contribution of spectral components varies while the intrinsic lags stay constant. From our spectral analysis, we conclude that the intrinsic QPO lags between soft and hard energy bands are truly changing on short time-scales of a few QPO cycles.

\subsection{Variations in the coronal geometry}
\label{sec:disc_coronalgeometry}

Because the flux-dependent variations in the QPO lags are unlikely to arise due to changes in the contribution of the different spectral components (disc and Comptonized emission) to the energy bands used, we investigate two physical mechanisms that can cause the lags to vary and the coherence to be decreased. First, we consider two different models for QPOs and QPO lags: the Lense-Thirring precession model as introduced by \citet{Ingram_2009LT} and the \texttt{vKompth} model as presented in \citet{Bellavita_2022}. In both models, changes in the coronal geometry on time-scales longer than the QPO time-scale may explain the variations in the QPO lags, and we suggest that such geometric changes also lead to a decreased coherence. Secondly, we discuss filtering effects on disc variability by the QPO mechanism, which may explain the lowered coherence, but is difficult to connect to the QPO lag - flux relation. 

From panel B) of Fig. \ref{fig:5panel} and the middle-right panel of Fig. \ref{fig:9panel}, we note that the measured lags between the very soft and hard bands at the QPO frequency are very large, up to 1 rad. Because the QPOs in \source{} are relatively weak, the measured QPO lags will be diluted by the broadband noise, which has much smaller lags in neighbouring frequencies. It is therefore reasonable to assume that the intrinsic disc--corona QPO lags may be on the order of $\pi$/2, or a quarter QPO cycle. Such a quarter-cycle delay between the maximum of the QPO signal in the disc band and the coronal band may be explained in the framework of a precessing hot flow causing the QPO. A good visualisation of such an effect is visible in Fig. 11 of \citet{Ingram_2016Fe}, Fig. 12 of \citet{Stevens_2016} and Fig. 3 of \citet{You_2020}, which show how the illumination of the inner accretion disc depends on the phase of a precessing hot flow corona. \citet{Stevens_2016} find with QPO phase-resolved spectroscopy of a type-B QPO in \sourceg{} that spectral parameters associated with the disc vary out of phase with coronal parameters by about 0.3 cycles, which can be explained by a precessing corona. From the point of view of the observer, the disc has a blueshifted side, especially for high-inclination sources like \source{} and \sourcem. If the corona at a given precession phase preferentially illuminates the blueshifted and hence Doppler-boosted side of the disc and a quarter of a precession cycle later shows a maximum (e.g. due to the maximum solid angle subtended) for the observer, the quarter-cycle delay between disc and corona is obtained naturally. 

Although the time-averaged lags shown in Fig. \ref{fig:9panel} may be consistent with the quarter-cycle delay in a precession model, Figs. \ref{fig:5panel}, \ref{fig:lagcoh_energy} and \ref{fig:6panel} show that the presence of the large (hard) lag varies on short time-scales. Here, we propose that the lags may be evidence for changes in coronal height. Reflection modelling has shown that the illumination pattern of the disc by the corona is strongly dependent on the vertical extent of the corona and small changes in coronal height may have a significant effect on the disc illumination \citep{Dauser_2014}. 

In a precession framework, the varying lags may arise if the coronal height changes significantly on short time-scales (slightly longer than the QPO time-scale). When the hard flux is low, the coronal height may be high and the corona illuminates up to fairly large radii. The blueshifted side of the illuminated disc could then show a peak in its emission about a quarter cycle before we see the peak in the emission from the corona, such that we measure large lags (e.g. flux bin 1 in Fig. \ref{fig:5panel}). When the hard X-ray flux is high, the implied coronal height is lower and the corona only illuminates the very inner parts of the disc. Due to a smaller illuminated area and stronger gravitational redshift at those small radii, the contribution to the variable spectrum of the part of the disc that is illuminated by the corona may then be much smaller and we do not measure the lag feature. This interpretation is in line with the conclusions in \citet{Bollemeijer_2024}, based on lags associated with reverberation. Given the importance of relativistic effects and the complicated interplay between different sources of variability, detailed modelling including non-linear variability is required to test the general idea outlined here.

% To explain the variable nature of the QPO lags in the precession framework, we assume that the illumination of the inner disc varies on time-scales longer than the QPO time-scale. 
We note that the changing lags themselves may lead to a reduction in the coherence at the QPO frequency, as such a non-stationary process will lead to non-unity coherence \citep{Nowak_1999}. Below the QPO frequency, the lags seem to be changing as well, but much less clearly than the QPO lags (see e.g. Fig. \ref{fig:5cohpanel}), and still the coherence is very low in low-frequency range. The changing coronal geometry may explain the low coherence on those time-scales, as it will lead to extra (low-frequency) variability in the corona that is not linearly related to disc variability. The fact that the coherence is near unity above the QPO frequency would imply that the QPO frequency is also the minimum time-scale on which the corona can significantly change shape, which is consistent with the Lense-Thirring precession model \citep{Fragile_2007,Ingram_2009LT}. We note that in \citet{Uttley_2025}, both the disc--power-law and power-law--power-law lags depend on the coronal geometry, so we would expect changes in both types of lags at non-QPO frequencies as well, if e.g. the coronal height changes significantly on time-scales of a few QPO cycles. We do not observe such behaviour in any of the sources we studied, although we note that the \citet{Uttley_2025} model assumes unity coherence, which is inconsistent with the data. Including non-linearities in the model may aid in explaining higher-order variability properties.

The geometry of the corona also determines the QPO lags in the \texttt{vKompth} model as presented by \citet{Bellavita_2022} and fitted to several sources such as GRS 1915+105, MAXI~J1535-571 and \source{} \citep{Garcia_2022,ZhangYuexin_2022,MaRuican_2023}. In the model, the energy-dependent lags arise due to Comptonization delays in the corona, which undergoes a quasi-periodic change in temperature. A larger corona will produce larger hard lags, because the light travel times inside the corona will increase. Soft lags at the QPO frequency are due to feedback, which means that a fraction of the coronal photons is reprocessed by the source of the seed photons, i.e. the accretion disc. Since the coronal size and feedback fraction are the main model parameters determining the amplitude and sign of the lags, at least one of those would have to change on the same short time-scales (tens of seconds) as the QPO lags themselves. Although fitting the \texttt{vKompth} model to different flux bins is beyond the scope of this work, we encourage model testing on results presented in this paper to learn more about the changes in the corona. Given the large difference in the phase lags between the lowest and highest flux bin, the corona will probably have to undergo significant changes in size or feedback fraction (or both). A change in feedback fraction affects the disc--power-law lags more than the lags between power-law bands, which is also what we observe here.

In any case, we expect that in both the precession and Comptonization interpretations of QPO lags, geometric changes on time-scales longer than the average QPO period are required to explain the observed behaviour. Again, such geometric changes also explain the decreased coherence observed below the QPO frequency. If both the lag feature and low coherence are related to changes in the coronal geometry, e.g. the vertical extent, it is no surprise that the effects are stronger for high inclination systems viewed more edge-on. A possible argument against large-scale geometric changes to explain the varying QPO lags, is that we would expect the reflection spectrum to show changes if the illumination pattern of the corona were to vary significantly. However, in the inset plot of Fig. \ref{fig:fluxresolved_spectra}, which shows the data/model ratio of the 3-10 keV range of the spectrum, we do not observe any changes to the Fe K line, implying that the reflection spectrum does not change strongly. These results can be reconciled if we assume that most of the emission creating Fe K line feature comes from the inner regions of the accretion disc, while the softest X-rays originate from farther out in the disc. Variations in the vertical extent may then affect the lags as the illumination pattern of the outer disc varies, while the reflection from the inner disc stays roughly constant. In the future, the combination of the spectral and timing response to changes in flux may constrain more detailed models.

\subsection{Filtering by the QPO}

The final explanation we consider for the low coherence between disc and coronal power-law bands, is that the QPO operates as either a low-pass or a high-pass filter. The low-pass filter scenario could potentially reproduce the data if there are two types of disc variability, e.g. in the accretion rate and the magnetic field, and we also assume that the magnetic field does not affect the disc emission, but only the coronal X-ray output. If the QPO filters out any magnetic field changes on time-scales shorter than the QPO frequency, those frequencies have high coherence (as is visible in Fig. \ref{fig:9panel}), while the QPO frequency and longer time-scales contain extra variability that is only present in harder energy bands. This extra variability due to e.g. magnetic field variations would thus lead to the reduced coherence and could also impact the coronal geometry, which may lead to the QPO lag -- flux relation, e.g. as also suggested in Section \ref{sec:disc_coronalgeometry}.

If the QPO acts as a high-pass filter, part of the variability from the disc, at time-scales longer than the QPO time-scale, does not reach the corona. The variability at longer time-scales is then partially incoherent between the disc and corona. Studying the energy-dependent power spectra of observations 108 and 145 in Fig. \ref{fig:9panel}, there may be extra variability in the (very) soft band at frequencies with a low coherence (i.e. below the QPO frequency), which is consistent with the idea of a filter that reduces the disc variability reaching the corona. However, the similar fractional rms for all energies at low frequencies, while the high frequency soft band variability is much smaller can also be explained by suppression of fast variability at larger disc radii, where the viscous time-scale is longer \citep{Churazov_2001,Mushtukov_2018}. Even if the QPO acts as a filter, it remains unclear what the underlying mechanism of the QPO is. We conclude that such a framework is worth further investigation, but doing so lies outside the scope of this work.
\newline

\begin{figure}
    \centering
    \includegraphics[width=\linewidth]{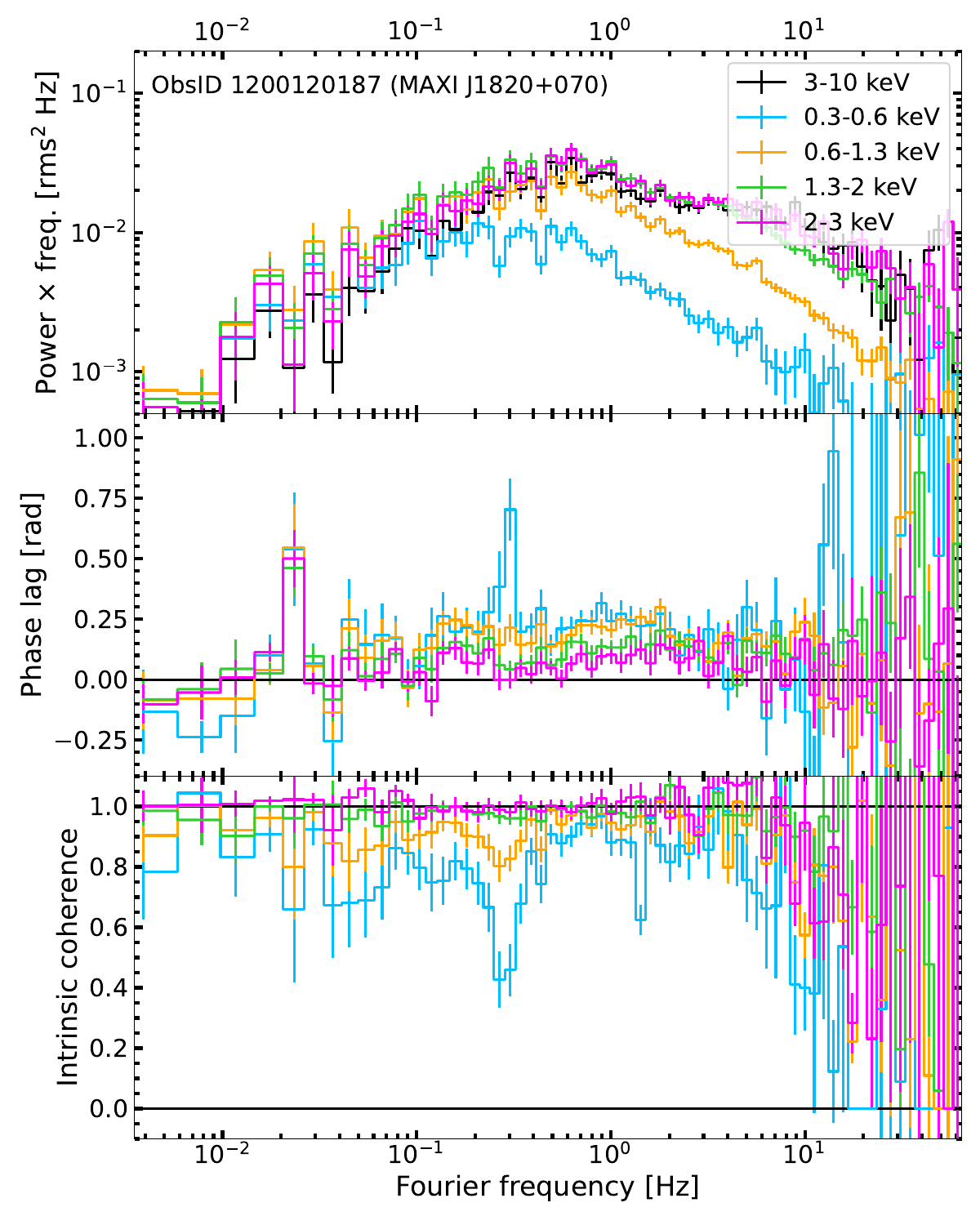}
    \caption{The power spectra, phase lag and coherence versus frequency spectra for \source{} observation 1200120187 during the bright decline of the outburst. The luminosity is a factor of three lower than during the peak of the hard state, and we observe no significant QPO in the power spectrum. However, around 0.28 Hz, where we expect the QPO to be based on the measured power-spectral hue, we observe a clear peak in the lags and a drop in coherence, reminiscent of spectral-timing properties in the soft-to-hard transition (see text).}
    \label{fig:pslagcoh_decline}
\end{figure}

\subsection{Luminosity-dependence of QPO spectral-timing properties}

We note that on longer time-scales, on the order of weeks as \source{} evolves during its outburst, the coronal geometry is thought to change based on evolution of the soft lags associated with reverberation \citep{Kara_2019,De_Marco_2021}. These longer-term changes may be related to the fact that we see a much stronger QPO lag -- flux relation for QPO frequencies below $\sim0.3$ Hz (Fig. \ref{fig:slopes_QPOf}). For example, the corona could be more vertically extended for low QPO frequencies, leading to larger changes in disc illumination as it precesses. 
\source{} shows more evolution of spectral-timing properties on long time-scales as well, especially as the luminosity drops. During the `bright decline', as defined by \citet{De_Marco_2021}, the QPOs become weaker and are often not significant in power spectra \citep{Stiele_2020}. In Fig. \ref{fig:slopes_QPOf}, we see that the empty symbols which correspond to the bright decline part of the outburst generally have shallower QPO lag -- flux slopes. In Fig. \ref{fig:pslagcoh_decline}, we show power spectra, phase lag and coherence versus frequency spectra of observation 1200120187 of \source, which has a factor $\sim3$ lower count rate than the brightest observations. Around 0.28 Hz, which is close to where we expect the QPO frequency to be based on the hue of $110\pm6$ deg (see table \ref{tab:obse_overview}), we see no QPO in the power spectrum, but there is a clear QPO-like feature in the lags and the coherence between the softest and hardest energy bands. The presence of the narrow feature in the lag and coherence, which is only visible when including very soft X-rays, is strongly reminiscent of the timing features recently observed in Cyg X-1 and during the soft-to-hard transition of several low-mass BHXRBs \citep{Koenig_2024,Mendez_2024,Fogantini_2025,Brigitte_2025}. \citet{Bellavita_2025} observed the cross-spectral feature in \source{} and identified it as an `imaginary QPO', as it is visible in the imaginary part of the cross-spectrum, but not in the power spectrum. At high luminosities, we observe a clear type-C QPO in the power spectrum, associated with large (and variable) hard lags and a drop in coherence at and below the QPO frequency between very soft and hard X-rays. During the bright decline, the QPO becomes weaker and disappears from the power spectrum, but we observe narrow lag and coherence features between very soft and hard X-rays at the expected QPO frequency, while the coherence recovers at lower frequencies. The latter spectral-timing properties have so far only been observed during the (lower luminosity) soft-to-hard transition. The fact that we measure similar spectral-timing features during the bright decline phase of the hard state may imply that the observational appearance of the underlying QPO mechanism depends strongly on the source luminosity during an outburst.

\section{Conclusions}
\label{sec:conclusions}

The results we obtained with the \nicer{} observations of \source{} in the hard state can be summarised as follows. 
\begin{enumerate}

    \item We studied the phase lags at the QPO frequency and how they depend on the instantaneous hard flux by binning light curve segments based on their hard flux. We find that the QPO lags between disc and power-law bands change significantly on time-scales of (tens of) seconds or a few QPO cycles. Lags between different coronal power-law energy bands show a much weaker or no dependence on the hard flux. The QPO lag -- flux relation is strongest for QPO frequencies below $\sim0.3$ Hz.
    \item When calculating the cross-spectral properties of disc and coronal power-law bands, we find that there is a clear hard lag peak at the QPO frequency, which is absent when measuring the lags between power-law bands. 
    \item The coherence between disc and coronal power-law bands is low at and below the QPO frequency and close to unity at higher frequencies, with the change closely following the QPO frequency. At the start of the hard state, when there are no QPOs, the coherence is close to unity across all \nicer{} energies, corroborating the link between the low coherence and the QPO. 
    \item The disc versus corona lag feature at the QPO frequency and the low coherence at and below the QPO frequency is observed in the high inclination sources \source{} and \sourcem{} (which also shows dips), but not in the low-inclination source \sourceg. This could imply that the observed behaviour is strongest for high-inclination BHXRBs. Due to the much lower brightness and smaller number of observations of both \sourcem{} and \sourceg, we could not assess whether the QPO lag -- flux relation is also present in these other sources.
    \item We discuss several scenarios to explain the observational results and conclude that our findings can possibly be explained by variations in the geometry of the corona on time-scales slightly longer than the QPO time-scale. The tentative inclination-dependence of the decreased low-frequency coherence suggests that the vertical extent of the corona changes most. The QPO lag feature could be due to reprocessing of coronal emission by the disc, which would only be visible when the corona is relatively `tall' and disappears when it is smaller. The lack of obvious variations in the reflection spectrum with hard flux suggests that the inner disc illumination does not vary much, but the outer parts do. The idea of a dynamic corona and its effects on spectral and timing properties requires investigation of sophisticated models with both broadband noise and QPO variability, in which the geometry of the corona is allowed to vary.
    \item On a time-scale of weeks, we find evidence for a dependence of the QPO amplitude and associated spectral-timing properties on the source luminosity during the outburst. At low luminosities, during the `bright decline' phase of the outburst of \source, the spectral-timing properties are reminiscent of those observed in the soft-to-hard transition in several accreting black hole. This implies that the observational appearance of the underlying QPO mechanism may depend on the luminosity.
\end{enumerate}

\section*{Acknowledgements}
We thank the referee, Adam Ingram, for valuable comments and suggestions that have improved the paper. N.B. and this work are supported by the research program Athena with project No. 184.034.002, which is (partially) financed by the Dutch Research Council (NWO). We made use of the overview provided by the BlackCAT Catalog of BHXRBs \citep{BlackCAT_2016}. This research makes use of the SciServer science platform (\url{www.sciserver.org}). SciServer is a collaborative research environment for large-scale data-driven science. It is being developed at, and administered by, the Institute for Data Intensive Engineering and Science at Johns Hopkins University. SciServer is funded by the National Science Foundation through the Data Infrastructure Building Blocks (DIBBs) program and others, as well as by the Alfred P. Sloan Foundation and the Gordon and Betty Moore Foundation.

%%%%%%%%%%%%%%%%%%%%%%%%%%%%%%%%%%%%%%%%%%%%%%%%%%
\section*{Data Availability}
This research has made use of data obtained through the High Energy Astrophysics Science Archive Research Center Online Service, provided by the NASA/Goddard Space Flight Center. The data underlying this article are available in HEASARC, at \url{https://heasarc.gsfc.nasa.gov/docs/archive.html}. 
Upon publication, a basic reproduction package for the results and figures presented in this paper will be available on Zenodo at \url{https://doi.org/10.5281/zenodo.14284350}.



\bibliographystyle{mnras}
\bibliography{biblio} % if your bibtex file is called example.bib% if your bibtex file is called example.bib

%%%%%%%%%%%%%%%%%%%%%%%%%%%%%%%%%%%%%%%%%%%%%%%%%%

%%%%%%%%%%%%%%%%% APPENDICES %%%%%%%%%%%%%%%%%%%%%

%%%%%%%%%%%%%%%%%%%%%%%%%%%%%%%%%%%%%%%%%%%%%%%%%%

\appendix
\section{Spectral-timing properties in \sourcem}
\label{app_J1803}

\begin{figure}
    \centering
    \includegraphics[width=\linewidth]{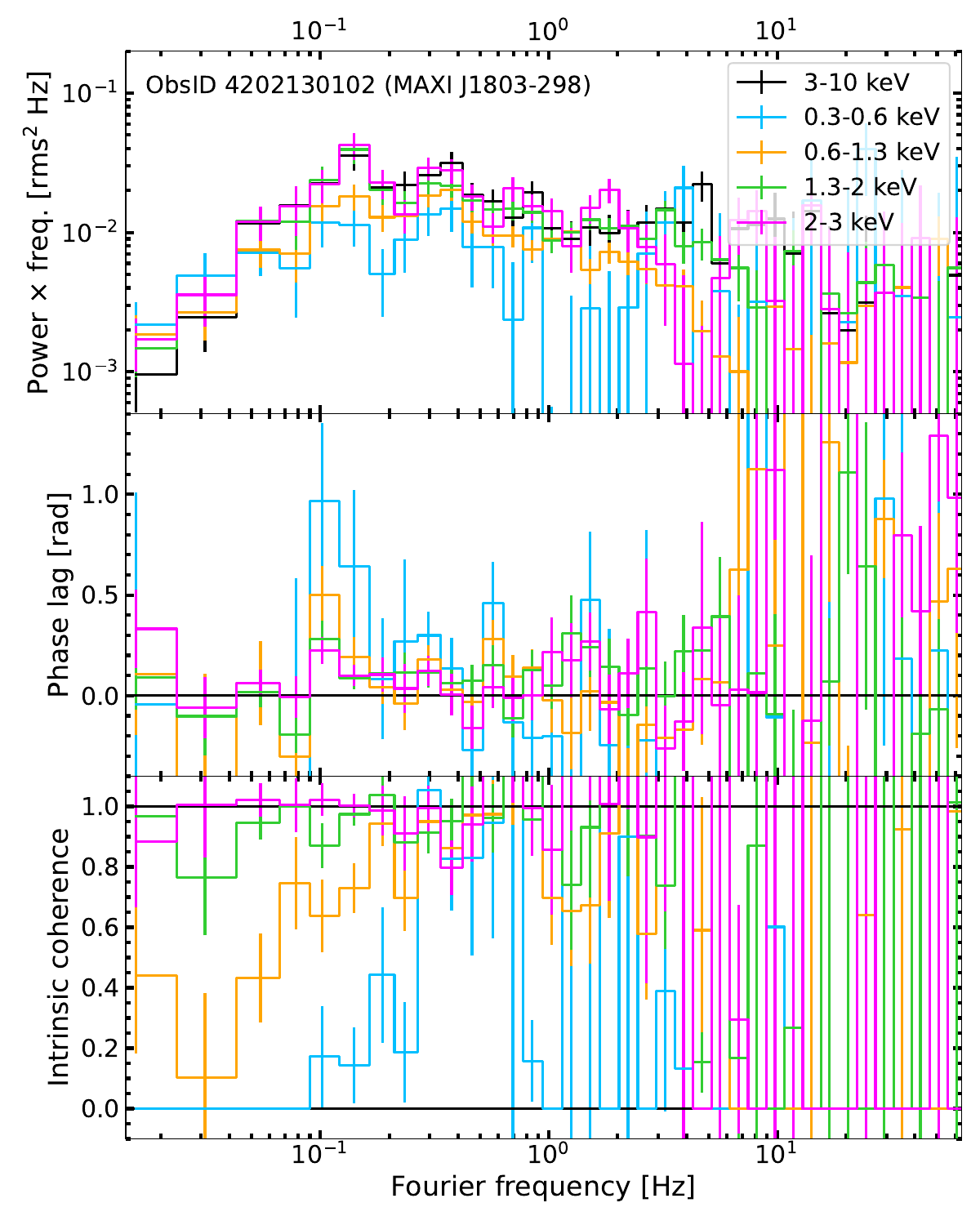}
    \caption{The power spectra and phase lag and coherence versus frequency spectra from observation 4202130102 of \sourcem. At the QPO frequency, around 0.13 Hz, a clear peak in the phase lags is visible and the coherence at and below the QPO frequency is low for the soft reference energies. The coherence pattern is similar to what we observe in Fig. \ref{fig:pslagcoh_1803}, but shifted to lower frequencies as the QPO frequency is lower here.}
    \label{fig:pslagcoh_1803_2}
\end{figure}

In Fig. \ref{fig:pslagcoh_1803_2}, we show the power spectra, lag- and coherence versus frequency spectra for ObsID 4202130102 from \sourcem. The QPO frequency is approximately 0.13 Hz. When using soft reference bands versus the hard band, the lag peak and low coherence at and below the QPO frequency are clearly at lower frequencies than for ObsID 4202130104 in Fig. \ref{fig:pslagcoh_1803}. We conclude that the lag- and coherence behaviour for disc and powerlaw bands follows the QPO frequency in \sourcem, similar to what we observe for \source.

% Don't change these lines
\bsp	% typesetting comment
\label{lastpage}
\end{document}